# A Polynomial-Time Deterministic Algorithm for an NP-Complete Problem

Xinwen Jiang* Holden Wool†

AUG, 2025

We introduce an NP-complete graph decision problem, the "*Multi-stage graph Simple Path*" (abbr. MSP) problem, which focuses on determining the existence of specific "global paths" in a graph $G$. We show that the MSP problem can be solved in polynomial ($O(|E|^{10})$) time, by proposing a polynomial-time graph algorithm and the proof of its correctness. Our result implies NP $=$ P. The algorithm leverages the data structure of *reachable-path edge-set* $R(e)$. By establishing the interplay between preceding decisions and subsequent decisions, the information computed for $R(e)$ (in a monotonically decreasing manner) carries all necessary contextual information, and can be utilized to summarize the "history" and to detect the "future" for searching "global paths". The relation of $R(e)$ of different stages in the multi-stage graph resembles the state-transition equation in dynamic programming, though it is much more convoluted. To avoid exponential complexity, paths are always treated as a collection of edge sets. Our proof of the algorithm is built upon a mathematical induction - based proving framework, which relies on a crucial structural property of the MSP problem: all MSP instances are arranged into the sequence $\{G_0, G_1, G_2, \ldots\}$, and each $G_j$ $(j > 0)$ in the sequence must have some $G_i$ $(0 \le i < j)$ that is completely consistent with $G_j$ on the existence of "global paths". As an auxiliary method, we have conducted tests using multiple AI systems. With the help of a suggested query list that covers the entire content of the paper, the paper has been verified by Doubao, DeepSeek, Kimi, iFlytek Spark, ERNIE Bot, Gemini, and GPT.

**Keywords and Phrases:** Graph algorithm, Computational complexity, Polynomial-time graph algorithm, NP-complete, MSP problem

## 1 INTRODUCTION

The research community has made great efforts (Wöginger, n.d.,2002) regarding the long-standing, well-known P vs. NP problem (Garey, & Johnson, 1979; Cook, 2003).

As categorized by Lance Fortnow (Fortnow, 2009; Fortnow, 2021), a range of techniques—e.g., diagonalization (Turing, 1936; Cantor, 1874; Baker, Gill, & Solovay, 1975; van Melkebeek, 2007), circuit complexity (Furst, Saxe, & Sipser, 1984; Razborov, 1985; Razborov, 1989; Razborov, & Rudich, 1997), proof complexity (Haken, 1985), and algebraic geometry (Mulmuley, & Sohoni, 2001; Bürgisser, & Ikenmeyer, 2011; Mulmuley, 2012)—have been adopted or proposed to prove NP $\neq$ P and other related problems.

Efforts to prove NP $=$ P have been mostly focused on searching for efficient algorithms for NP-complete problems. A succession of successful algorithms for hard problems (e.g., the AKS algorithm for Primality Test (Agrawal, Kayal, & Saxena, 2004), the quasi-polynomial-time algorithm for Graph Isomorphism (Babai, 2016), the holographic algorithm for counting problems (Valiant, 2002) and the many constructive disproofs of prominent conjectures in cryptography (Viola, 2018)) have repeatedly shown that people have grossly underestimated the reach of efficient computation across diverse contexts and thus inspiring such endeavors. Don Knuth (Knuth, 2002) believes that NP = P, yet contends that even if a proof were given, it might lack constructiveness; or, even if an algorithm were found, it would be excessively complex to hold practical significance.

---

* Corresponding author. Supported by the National Natural Science Foundation of China ("Research on the Complexity to Solve An NPC Problem", No. 61272010). Email: xinwenjiang@sina.com; xwjiang@nudt.edu.cn; xinwenjiang@xtu.edu.cn
† Email: holdenwool@foxmail.com

First introduced in Jiang, Peng, and Wang (2010), the "*Multi-stage graph Simple Path*" (abbr. MSP) problem was shown to be polynomial-time reducible from the well-known NP-complete *Hamilton Circuit* (abbr. HC) problem. Ten years later, a Chinese-language paper (Jiang, 2020) was published in July 2020, in which a polynomial-time algorithm for the MSP problem was presented. This has caused widespread concerns and extensive discussions.

This paper focuses directly on a polynomial-time graph algorithm (the *ZH algorithm*) for the NP-complete MSP problem, which further greatly simplifies and refines the proof given in Jiang (2020). For sake of being self-contained, of the problems caused by different languages and of the convenience of reading, we will include the formal definitions of the MSP problem and the ZH algorithm as specified in Jiang (2020). The current paper is a significantly simplified and refined proving framework of the correctness of the ZH algorithm together with a rigorous and complete proof:

1. Simplification of the induction variable "$f(G)$". The right-hand addition operand "$L$" is omitted from the original "$f(G) = \left(\sum_{v \in V-\{S,D\}} (d^-(v) - 1)\right) + L$", so that our mathematical induction on "$f(G)$" can be done by a "split" transformation for reducing "$\sum_{v \in V-\{S,D\}} (d^-(v) - 1)$", without further "compact" transformation for reducing "$L$" (as required previously).
2. Restriction to a more specific problem called $2-$MSP. In-degrees are bounded by 2 and out-degrees never exceeds in-degrees. Hence, during the split transformation, the indeterminate discussion of "$x + y = z$" ($x, y, z \in \{1,2,\ldots\}$) simplifies to the precise case of "$1 + 1 = 2$", forming a clear "either-or" logical structure.
3. Analytical definitions of basic operators and justification of their consistency with the original procedure definitions. Procedure forms facilitate analyzing and reducing computational complexity, which is the ultimate goal of the paper; while analytical forms better describe the mathematical properties of basic operators.

The insight behind our focus on MSP lies in discovering a rich structural property of the problem, which naturally gives rise to a mathematical induction-based proving framework. Subsequently, designing a polynomial-time algorithm that fits this framework becomes our pursuit. Meanwhile, that property and the proving framework also make it possible for the rigorous proof of the correctness of the algorithm.

The online materials of the paper are available at:

https://tcsrepositories.github.io/PvsNP/,

https://weibo.com/p/1005051423845304.

## 2 THE PROBLEMS OF MSP, $2-$MSP

***Definition 1 (Labeled multi-stage graph).*** *A labeled multi-stage graph* $G = < V, E, S, D, L, \lambda >$ *is a special directed acyclic graph (DAG), where:*

1. $V$ is the set of vertices, which is divided into $L+1$ ($L \in \mathbb{N}$) *stages*: $V = \bigcup_{0 \le l \le L} V_l$ ($V_i \cap V_j = \emptyset, 0 \le i \le L, 0 \le j \le L, i \ne j$).[1] $u$ is *a vertex of stage* $l$, if $u \in V_l$ ($0 \le l \le L$).
2. $V_0$ contains the single *source* $S$. $V_L$ contains the single *sink* $D$.
3. $E$ is the set of edges. For the convenience of algorithmic processing and complexity analyzing, each edge is denoted by $\langle u, v, l \rangle$ ($u \in V_{l-1}, v \in V_l, 1 \le l \le L$), which is called *an edge of stage* $l$. We use $d^-(v)$ and $d^+(v)$ to each denote the in-degree and out-degree of $v$. A path $P \subseteq E$ (directed by default) from vertex $a$ to $b$ is denoted by $a - \cdots - b$.
4. $\lambda$ is a mapping from $V - \{S\}$ to $2^E$. $\lambda(v)$ ($v \in V - \{S\}, \lambda(v) \subseteq E$) is called the *label of* $v$.

***Definition 2 (ω-path, σ-path).*** *Let $G =< V, E, S, D, L, \lambda >$ be a labeled multi-stage graph. (1) If $P = a - \cdots - b \subseteq E$ such that $P' \subseteq \lambda(v)$ for each $P' = a - \cdots - v \subseteq P$, then $P$ is called a weak simple path (abbr. $\omega$-path). (2) If $P = S - \cdots - D \subseteq E$ such that $P' \subseteq \lambda(v)$ for each $P' = S - \cdots - v \subseteq P$, then $P$ is called a simple path (abbr. $\sigma$-path).*

The above definition should be distinguished from the conventional concept of "simple path" in graph theory. The latter only requires the path to traverse a vertex no more than once, which is always satisfied in a DAG. However, edges on a path might be rejected by labels on the path, to describe which we borrow the term "simple".

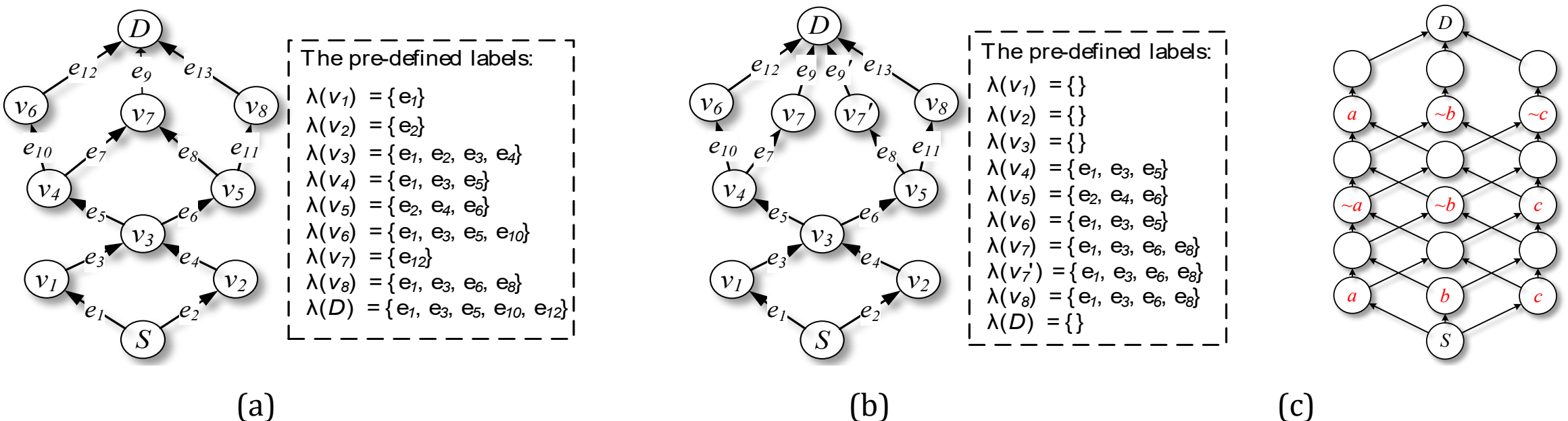

Figure 1: Labeled multi-stage graphs

***Definition 3(a) (The "Multi-stage graph Simple Path" (MSP) problem).*** *The MSP problem asks whether a given labeled multi-stage graph $G =< V, E, S, D, L, \lambda >$ contains a $\sigma$-path.*

The MSP instance illustrated in Figure 1(a) contains $\sigma$-paths (e.g., $S - v_1 - v_3 - v_4 - v_6 - D$) and $\omega$-paths (e.g., $v_1 - v_3 - v_4, S - v_2 - v_3 - v_5$), while the one in Figure 1(b) contains no $\sigma$-path. The existence of $\sigma$-paths in a graph depends on its structure, as well as its labels.

For technical reasons, we will further focus on a restricted form of MSP.

***Definition 3(b) (2 − MSP).*** *The $2 - MSP$ problem is a special MSP problem fulfilling:*

1. $d^+(v) > 0$ ($v \in V - \{D\}$); $d^-(v) > 0$ ($v \in V - \{S\}$). *(Each vertex should appear on some path $S - \cdots - D \subseteq E$.)*
2. $d^-(v) \le 2$ ($v \in V - \{S, D\}$); $d^-(v) = 1$ ($v \in V_{L-1}$). *(In-degrees are limited.)*

[1] Indices are in ℕ (the set of nature numbers, including zero) by default.

3. $(\forall v \in \bigcup_{1<i<L} V_i)\left((d^-(v) \leq 1) \Rightarrow \left(\substack{\forall \langle a,b,h \rangle \in \\ (v - \cdots - D) \subseteq E}\right)(d^-(a) \leq 1)\right)$. *(Roughly, if a vertex is not multi-in-degree, then neither is any vertex except $D$ on subsequent paths.)*
4. $(\forall v \in \bigcup_{1<i<L-2} V_i)(d^+(v) \leq d^-(v))$.
5. $L \geq 5$ and $\lambda(D) = E$.

The basic structure of $2-\mathrm{MSP}$ is shown in Figure 1(c), generally: for each vertex, its in-degree and out-degree are within 2; for each vertex of stage $L-1$, its in-degree equals 1. In the inductive proof of our algorithm, we have to construct a pair of logically equivalent graphs. We list several key structural properties item by item in the definition of $2-\mathrm{MSP}$, so as to check the properties of the constructed graph against the original graph one by one, and to also facilitate the adaptation of the changes inevitably caused by us to the constructed graph.

***Theorem 1 (NP-completeness).*** $2-MSP \in NPC$. *(trivial; proof can be found in Appendix A.1)*

The problems of MSP and $2-\mathrm{MSP}$ properly provide a "split"-based inductive proving framework towards the resolution of the P vs. NP problem.

## 3 THE ZH ALGORITHM FOR $2-\mathrm{MSP}$

### 3.1 Basic operators

The ZH algorithm utilizes four basic operators on edge sets for a given $G =< V, E, S, D, L, \lambda >$, as follows.

***Operator 1 ($[ES]_u^v$).*** *Given $ES \subseteq E, \{u,v\} \subseteq V$. $[ES]_u^v =_{def} \{e | e$ is on some path $u - \cdots - v \subseteq ES\}$.*

When discussing connectivity, such paths "$u - \cdots - v$" are taken as a whole set of edges (via polynomial-time connectivity check), rather than being distinguished from each other (via exponential-time path enumeration). The same is with the below.

Operator 1 extracts the edges of paths between two designated vertices for a certain collection of edges. The CONNECTIVITY problem is known to be solvable in $O(|E|)$ ($|E|$ denotes the cardinality of the set $E$), hence Operator 1 can be done in $O(|E|)$.

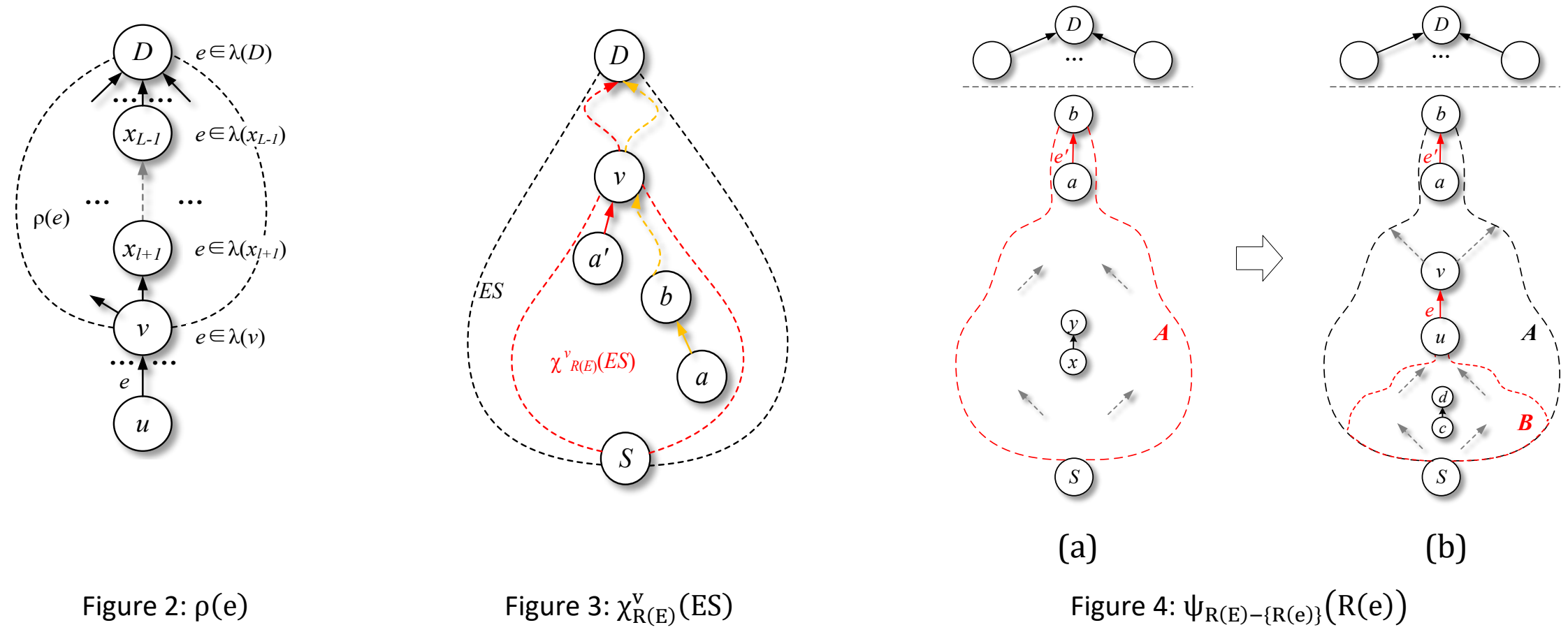

Figure 2: $\rho(e)$    Figure 3: $\chi^v_{R(E)}(ES)$    Figure 4: $\psi_{R(E)-\{R(e)\}}(R(e))$

***Operator 2 ($\rho(e)$).*** *Given $e = \langle u,v,l \rangle \in E$. $\rho(e) =_{def} [\{\langle a,b,k \rangle \in E \mid e$ belongs to both $\lambda(a), \lambda(b)\}]_v^D$.*

The operator captures a necessary condition of $\sigma$-path existence. By definition, $\rho(e)$ collects the edges on every $v - x_{l+1} - \cdots - x_{L-1} - D \subseteq E$, if $e \in \lambda(v) \cap \lambda(x_{l+1}) \cap \cdots \cap \lambda(x_{L-1}) \cap \lambda(D)$, as illustrated by the region enclosed by dotted curves in Figure 2. As can be hence observed, for each $\langle x_{i-1}, x_i, i \rangle$ $(1 \leq i \leq L)$ on a $\sigma$-path $P = x_0 - x_1 - \cdots - x_L$ $(x_0 = S, x_L = D)$, we have $\langle x_{i-1}, x_i, i \rangle \in \bigcap_{i \leq j \leq L} \lambda(x_j)$ and thus $\rho(\langle x_{i-1}, x_i, i \rangle) \supseteq [P]_{x_i}^{D}$.

The cost of $\rho(e)$ can be $O(|E|)$.

We use the notation $R(e)$ (i.e., $R(\langle u, v, l \rangle)$) as a global variable, which initially holds the result of $\rho(e)$ and will be updated later by the ZH algorithm. $R(e)$ carries all contextual information needed by $e$ to detect the "future" (i.e., the containment of $e$ by labels) for searching $\sigma$-paths. Let's denote $R(E) = \{R(e) | \ e \in E\}$.

***Definition 4 (ρ-path, reachability).*** *Each path $v - \cdots - D \subseteq R(e)$ $(e = \langle u, v, l \rangle \in E)$ is called a reachable path (abbr. ρ-path) of e. $R(e)$ is called the ρ-path edge-set of e, which characterizes the reachability of e during the computation of the ZH algorithm. $R(E)$ is called the collection of ρ-path edge-sets.*

***Operator 3 ($\chi_{R(E)}^{v}(ES)$, procedural form).*** *Given $ES \subseteq E$, $v \in V_l$ and the collection of ρ-path edge-sets $R(E)$. The result of Operator 3, given as the following "procedural form", equals to the final stable $ES'$:*

(1) $ES' \leftarrow ES$

(2) **for** $e = \langle a, b, k \rangle \in ES'$

  **if** $[R(e) \cap ES']_b^v = \emptyset$ $(k < l)$[2]

    **then** $ES' \leftarrow ES' - \{e\}$

  **if** $[R(e)]_v^D = \emptyset$ $(k = l \neq L)$

    **then** $ES' \leftarrow ES' - \{e\}$

(3) $ES' \leftarrow [ES']_S^v$

(4) **repeat** (2),(3) **until** $ES'$ becomes stable

The operator utilizes the $\rho$-path edge sets in $R(E)$ to compact $ES$, as illustrated by the innermost region enclosed by dotted curves in Figure 3. Intuitively speaking, the compacted set $ES'$ is the collection of connected edges $e = \langle a, b, k \rangle \in ES$, such that some $\rho$-path $P \subseteq R(e)$ should "cling" onto the edges in the compacted set to "climb" towards $v$ (i.e., $[P]_b^v \subseteq ES'$).

It should be noted that, $b$ can be $v$ or $D$. To maintain the intended semantics of the operator, for these boundary conditions, the definition is slightly different. When $b = v$ $(v \neq D)$, the pruning of $e$ is decided on the content of $[R(e)]_v^D$, instead of $[R(e) \cap ES']_b^v$. When $b = v$ $(v = D)$, we always have $R(e) = \emptyset$ and thus we shall never prune $e$ simply by the content of $R(e)$.

The result of Operator 3 is uniquely determined, regardless of the order of choice of the edges to be pruned during the iteration (see Theorem 2 in Appendix A.2).

Step (2),(3) can be done in $O(|E|^3)$. The execution can terminate within $|E|$ iterations, since at least one edge is pruned per round. Thus, the overall cost is $O(|E|^4)$.

---

[2] $\emptyset$,{} are not distinguished in the paper. The result of set operations can be united with {} to avoid null reference.

***Operator 4** ($\boldsymbol{\psi_{R(E)-\{R(e)\}}(R(e))}$**, procedural form).** Given $e = \langle u, v, l\rangle \in E$ ($1 < l < L$) and the collection of $\rho$-path edge-sets $R(E)$. Operator 4 uses $R(E) - \{R(e)\}$ to restrain $R(e)$, given as the following "procedural form":*

(1) **for** $e' = \langle a, b, k\rangle \in R(e)$ (from $k = l + 1$ to $k = L$)

$$\boldsymbol{A} \leftarrow \chi^b_{R(E)}\left(\{\langle x, y, i\rangle \in E \,|\, e' \in [R(\langle x, y, i\rangle) \cap \lambda(b)]^b_y\} \cup \{e'\}\right)$$

$$\boldsymbol{B} \leftarrow \chi^u_{R(E)}\left(\{\langle c, d, j\rangle \in \boldsymbol{A} \,|\, \{e, e'\} \subseteq [R(\langle c, d, j\rangle) \cap \boldsymbol{A}]^b_d\}\right)$$

**if** $\boldsymbol{B} = \emptyset$

**then** $R(e) \leftarrow R(e) - \{e'\}$

(2) $R(e) \leftarrow [R(e)]^D_v$

(3) **repeat** (1),(2) **until** $R(e)$ becomes stable

The result of Operator 4 is uniquely determined, regardless of the order of choice of the edges to be pruned during the iteration (see the following Theorem 3).

Note that, the $R(e)$ modified by Operator 4 now becomes a subset of the original $R(e)$, but we would rather still call each $v - \cdots - D \subseteq R(e)$ a *$\rho$-path of e* and call $R(e)$ *the $\rho$-path edge-set of e*. $\psi_{R(E)-\{R(e)\}}(R(e))$ utilizes $(R(E) - \{R(e)\})$ to restrict each $e' \in R(e)$, thus "binding" related $\rho$-path edge-sets all together.

It will be seen later that, Operator 4 is going to be used iteratively by the ZH algorithm to prune $R(e) \in R(E)$, until each $R(e) \in R(E)$ becomes stable; the computation is always strait forward and decreases monotonically. This technique lies in the center of the ZH algorithm, which realizes the exploitation of the relation between local strategies and global strategies. This resembles the paradigm of dynamic programming, nevertheless much more convoluted.

The constraint imposed on each $\langle a, b, k\rangle \in R(\langle u, v, l\rangle)$ by Operator 4 arises from two compacting operations—each targeting either $\langle u, v, l\rangle$ or $\langle a, b, k\rangle$:

- For $\langle a, b, k\rangle$, the compacted set $\boldsymbol{A}$ is a subset of $\lambda(b)$. Each $e'' \in \boldsymbol{A}$ ($e'' \neq \langle a, b, k\rangle$) eventually "falls" into $\lambda(b)$ by "walking" along a path that traverses $\langle a, b, k\rangle$, i.e., $R(e'')$ contains a $\rho$-path traversing $\langle a, b, k\rangle$. We can imagine $\boldsymbol{A}$ as a "gourd" hanging under the "handle" $\langle a, b, k\rangle$, as depicted in Figure 4(a).
- For $\langle u, v, l\rangle$, the set $\boldsymbol{B}$ is compacted from the set $\boldsymbol{C} =$ $\{\langle c, d, j\rangle \in \boldsymbol{A} \,|\, \{\langle u, v, l\rangle, \langle a, b, k\rangle\} \subseteq [R(\langle c, d, j\rangle) \cap \boldsymbol{A}]^b_d\} \subseteq \boldsymbol{A}$. Regarding $\langle u, v, l\rangle$ as a "handle", then obviously $\boldsymbol{B} = \chi^u_{R(E)}(C) \subseteq \boldsymbol{A}$ and $\boldsymbol{B}$ is also like a "gourd" hanging under the "handle" $\langle u, v, l\rangle$, as depicted in Figure 4(b).
- Intuitively speaking, if $\langle a, b, k\rangle$ is kept in $R(\langle u, v, l\rangle)$, there must exist $P = S - \cdots - u \subseteq E$ such that $\{\langle u, v, l\rangle, \langle a, b, k\rangle\} \subseteq R(e''')$ for each $e''' \in P$. Meanwhile, all those paths like $P$ must fulfill the strict constraint that: suppose all the edges on those paths form a set $ES$, then $\chi^u_{R(E)}(ES) \neq \emptyset$.

It should also be noted that, the result of Operator 4 does not depend on the order of edge choice during the iteration (see Theorem 3 in Appendix A.2).

The sets $\boldsymbol{A}$ and $\boldsymbol{B}$ can be computed within $|E| * O(|E|^4)$ and therefore step (1) can be finished in $|E| * |E| * O(|E|^4)$. The execution will terminate before it reaches $|E|$ iterations, since at least one edge is pruned per round. Overall, the cost is $|E| * |E| * |E| * O(|E|^4) = O(|E|^7)$.

### 3.2 The ZH algorithm, the temporal cost and the necessity proof

With the above basic operators, the ZH algorithm can be henceforth given in the following Algorithm 1. Detailed motivations of the algorithm are discussed in Section 4.

The edges contained in $R(e)$ in step 1 are initially computed by Operator 2 and denoted by $R_0(e) =_{\text{def}} \rho(e)$, $R_0(E) =_{\text{def}} \{R_0(e)|e \in E\}$. $R(e)$ is pruned thereafter with $|R(e)| \leq |E|$ decreasing monotonically, until this procedure eventually stops.

In step 2, Operator 4 leverages $(R(E) - \{R(e)\})$ to restrict each $e' \in R(e)$ for the determination of $\sigma$-path existence, by not only detecting the "future" (using the $\rho$-paths in $R(e)$) but also summarizing the "history" (by binding $(R(E) - \{R(e)\})$ with $R(e)$).

---

ALGORITHM 1: The ZH Algorithm

---

Input: $G =< V, E, S, D, L, \lambda >$ in $2 - \text{MSP}$

Output: 'yes' or 'no' decision on $\sigma$-path existence

1. $R(E) \leftarrow \{R(e)|R(e) \leftarrow \rho(e), e \in E\}$

2. **for** $e = \langle u, v, l\rangle \in E$ $(2 \leq l < L)$

    call $\psi_{R(E)-\{R(e)\}}\big(R(e)\big)$ to prune $R(e) \in R(E)$

3. **repeat** step 2 **until** each $R(e) \in R(E)$ becomes stable

4. $G$ contains a $\sigma$-path iff. $\chi^{D}_{R(E)}(\lambda(D)) \neq \emptyset$

---

Supplementary video demos and running instances (on $\text{K} - \text{SAT}$) of the ZH algorithm are provided in the online materials. As also discussed therein, the algorithm has been validated on a wide range of test cases, including a large number of hard $3 - \text{SAT}$ instances of moderate sizes generated by a phase-transition-theory based model (Xu, & Li, 2000; Xu, Boussemart, Hemery, & Lecoutre, 2007).

***Definition 5 (The compact kernel).*** *The resulted $\chi^{D}_{R(E)}(\lambda(D))$ in step 4 of the ZH algorithm is called the compact kernel of $G$.*

Our result is amazingly simple, which is given as the following conjecture:

***Conjecture 1 (The Compact Kernel Conjecture).*** *G contains a $\sigma$-path iff. the compact kernel of G is not empty.*

***Theorem 4 (The cost).*** *The cost of the ZH algorithm can be $O(|E|^{10})$. (Proof see Appendix A.3)*

***Theorem 5 (The necessity).*** *If G contains a $\sigma$-path, then the compact kernel of G is not empty. (the direction of necessity is naturally trivial; proof see Appendix A.4)*

### 3.3 The sufficiency proof

Before the proof of sufficiency, two notations, one metric and a specially constructed algorithm need to be defined.

***Definition 6 ($ES[i:j]$).*** *Let $ES \subseteq E$. $ES[i:j]$ denotes the set of all edges of stages from $i$ to $j$ in $ES$, where $1 \le i \le j \le L$. If $i > j$, $ES[i:j] = \emptyset$.*

***Definition 7 ($ZH\backslash step4$).*** *$ZH\backslash step4$ stands for all the steps of the $ZH$ algorithm except step 4.*

To apply mathematical induction, the following metric for $G$ is required.

***Metric 1.*** $\boldsymbol{f(G) = \sum_{v \in V-\{S,D\}}(d^-(v) - 1)}$.

The ZH algorithm is then embedded in a **Proving algorithm** (abbr. PA, see Algorithm 2), which is specially constructed to set up the sufficiency proof.

The edge set $ES1_{sub}$ in the PA fulfills the five criteria. Here is the explanation:

- The composition of $ES1_{sub}$ is defined by criterion (i), where the subset $ES2 \subseteq ES1[L:L]$ is used to control the last stage; and then for each $\langle y, D, L\rangle \in ES2$, the subset $\lambda_{sub}(y) \subseteq \lambda(y)$ is used to constitute the rest of $ES1_{sub}$.
- The definition of $\lambda_{sub}(y)$ by criterion (ii) is based on a key insight that each pair of edges "$\langle y, D, L\rangle, e$" involved in the computed relation "$\langle y, D, L\rangle \in R(e)$" by the ZH algorithm must be on some $\sigma$-path. And every $\sigma$-path that can traverse $y$ can be kept (except $\langle y, D, L\rangle$) in $\lambda_{sub}(y)$.
- Criterion (iii) defines the relation between $\lambda_{sub}(y)$ and $ES1_{sub}$. Since $P = S - \cdots - y - D$ is a $\sigma$-path, $P[1:L-1] \subseteq \lambda(y)[1:1] \cup \{e \in \lambda(y)[2:L] | \langle y, D, L\rangle \in R(e)\}$. Generally, $\lambda_{sub}(y) \subseteq \lambda(y)[1:1] \cup \{e \in \lambda(y)[2:L] | \langle y, D, L\rangle \in R(e)\}$. If $P \subseteq ES1_{sub}$, $\lambda_{sub}(y)$ is demanded to contain the whole $P[1:L-1]$.
- By criterion (iv), $[ES1_{sub}]_S^D[1:L-1]$ should be a tree.
- Criterion (v) defines the relationship between different $\lambda_{sub}(y)$.

Step 1 of the PA is actually the ZH algorithm. The PA only makes a sufficient judgment. The PA first works according to the ZH algorithm, and then determines the existence of some $\sigma$-path (the *solution* found by the PA) contained by the given edge set $ES1_{sub} \subseteq ES1$. $ES1_{sub}$ should meet specific strict conditions, but such kind a $ES1_{sub}$ might not exist. The inference made by the PA is, if $ES1_{sub}$ exists, then $ES1_{sub}$ contains the demanded $\sigma$-path and every edge in$[ES1_{sub}]_S^D[L:L]$ is on such a $\sigma$-path.

---

ALGORITHM 2: The Proving Algorithm

---

Input: $G =< V, E, S, D, L, \lambda >$ in $2-\mathrm{MSP}$

Output: 'yes' or 'no' decision on $\sigma$-path existence in step 4 of the PA

1. apply $ZH\backslash step4$ on $G$ to generate $R_0(E)$ and the stable $R(E)$

2. $ES1 \leftarrow \chi_{R(E)}^D(\lambda(D))$

3. **if** $[ES1_{sub}]_S^D[L:L] = ES2$ (where $ES1_{sub}$ should obey the criteria (i),(ii),(iii),(iv),(v))

**then** $(\forall \langle w, D, L\rangle \in [ES1_{sub}]_S^D)(\exists\ \sigma-\text{path}\ S - \cdots - w - D \subseteq ES1_{sub})$

The criteria on the constitution of $ES1_{sub}$ are as follows:

(i) $ES1_{sub} = ES2 \cup \left(\left(\bigcup_{\langle y,D,L\rangle \in ES2} \lambda_{sub}(y)\right) \cap ES1\right)$, where $\emptyset \ne ES2 \subseteq ES1[L:L]$.

(ii) $\lambda_{sub}(y) \subseteq \lambda(y)[1:1] \cup \{e \in \lambda(y)[2:L] | \langle y, D, L\rangle \in R(e)\}$, where $\langle y, D, L\rangle \in E$.

**(iii)** If a $\sigma$-path $P = S - \cdots - y - D \subseteq ES1_{sub}$, then $P[1{:}L-1] \subseteq \lambda_{sub}(y)$.

**(iv)** $ES1_{sub}$ contains one of $\langle *, v, k \rangle$ at most for each $v \in V_k$ $(1 < k < L)$.

**(v)** If $|ES2| > 1$, there exists a path $S - a_1 - \cdots - a_i \subseteq [ES1_{sub}]_S^D$ $(i > 1)$ such that $[ES1_{sub}]_S^D$ contains one $\langle a_j, *, j+1 \rangle$ at most for each $a_j$ $(1 \leq j < i)$ while two $\langle a_i, *, i+1 \rangle$ at least, and $S - a_1 - \cdots - a_i \nsubseteq \lambda_{sub}(y)$ for $\langle y, D, L \rangle \in ES2$.

---

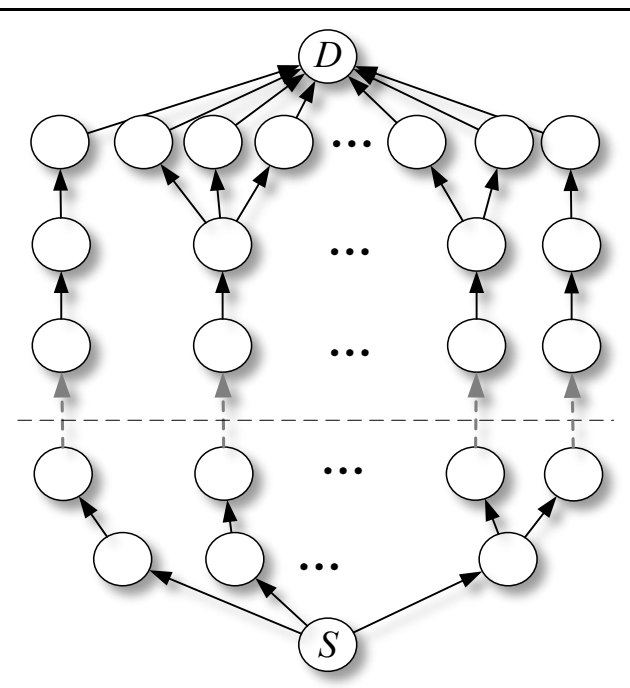

Figure 5: Illustration of Lemma 1

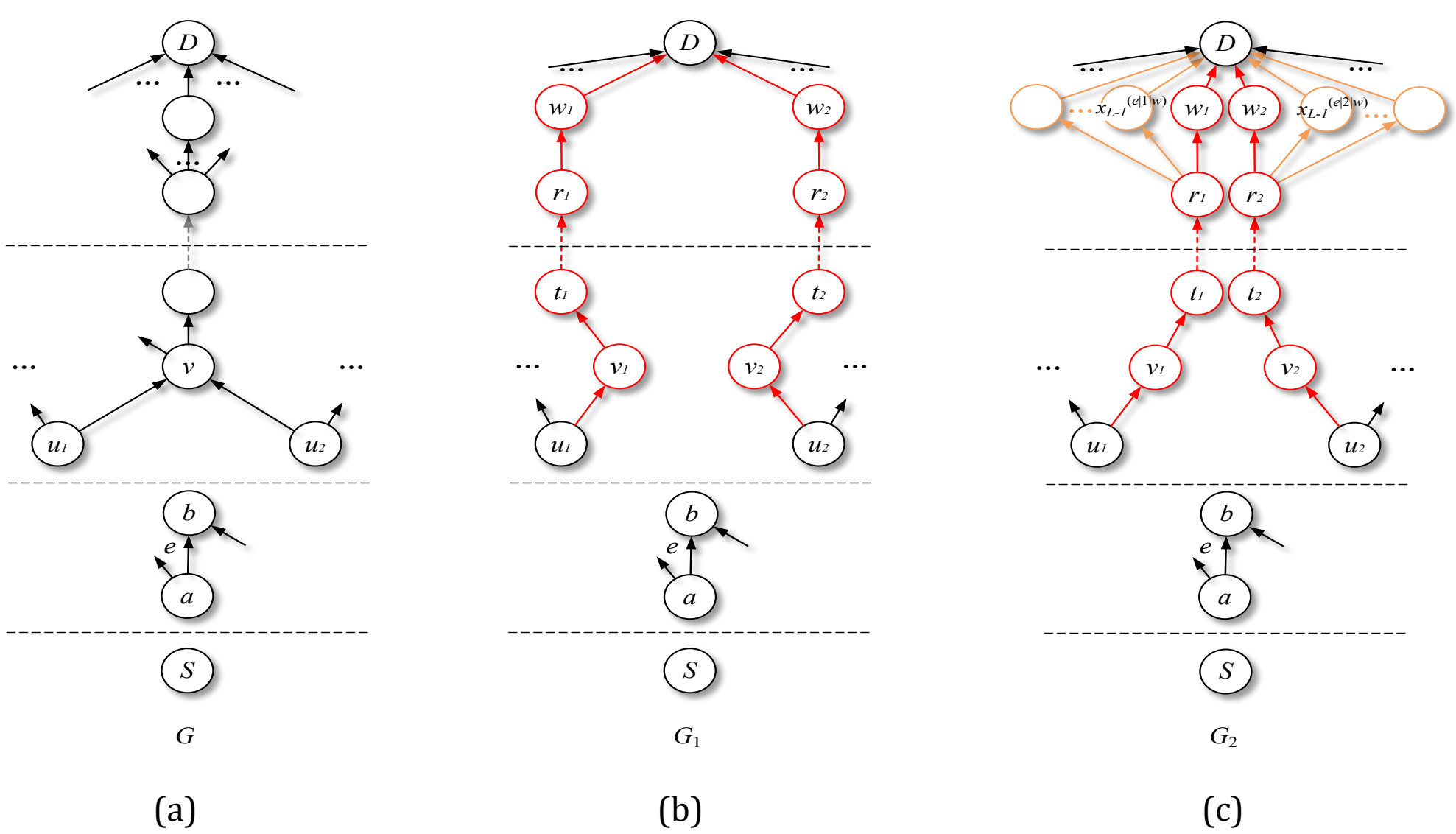

Figure 6: Illustration of Lemma 2 ($v \in V_l$, $1 < l < L$; recall $L \geq 5$ by Definition 1)

***The proving framework of mathematical induction on $f(G)$.*** *By Definition 3(b) (item 1), we have $f(G) \geq 0$ when applying Metric 1. For any G, if $f(G) = 0$, it can be proved that the PA can make a correct decision (see the following Lemma 1). Assuming that the PA can make a correct decision for any $G'$ that $f(G') < m$ $(m > 0)$ (**H1**), we can prove that the PA can make a correct decision for any G that $f(G) = m$ (see the following Lemma 2).*

***Definition 8 ($(*$ of $G)$).*** *To specify the context when necessary—i.e., a given property in G and its peer in another graph say $G'$—we use the indicators $(*$ of $G)$ and $(*$ of $G')$ respectively. If more than one "$(*$ of $G)$" are intended* *for a bundle of attributes in the same graph, just use one outermost "$(*\ of\ G)$" for brevity. For instance, each* $ES1, R(E)$ *within* $\left(\chi^{D}_{R(E)}(ES1)\ of\ G'\right)$ *actually refers to* $(ES1\ of\ G'), (R(E')\ of\ G')$.

The major difficulty and challenge of the above mathematical induction-based proof is that, given the set $ES1_{sub}$ for the input $G$, we shall construct a mathematically equivalent new $ES1_{sub}'$ for some other graph $G'$ that is "smaller" than $G$. Indeed, some parallels can be drawn (as will be revealed during the proof) on the logical power of construction between the $ES1_{sub}'$ by our approach and the uncomputatble functions (Church, 1936; Turing, 1936) by diagonalization.

***Lemma 1.*** *Let* $G =< V, E, S, D, L, \lambda >$ *be the input to the* $PA$ *and there is no multi-in-degree vertex at stage* $1,2, \dots, L-1$ *in* $G$ *(see Figure 5). After applying the* $PA$ *on* $G$*, if* $[ES1_{sub}]^{D}_{S}[L{:}L] = ES2$*, then* $(\forall \langle w, D, L\rangle \in [ES1_{sub}]^{D}_{S})(\exists\ \sigma - path\ S - \cdots - w - D \subseteq ES1_{sub})$*. (proof see Appendix A.5)*

***Lemma 2.*** *Given the mathematical induction hypothesis H1 that the* $PA$ *can make a correct decision for any* $G'$ *that* $f(G') < m\ (m > 0)$*. Let* $G =< V, E, S, D, L, \lambda >$ *be the input to the PA,* $f(G) = m$*, the vertex* $v$ *of stage* $l$ $(1 < l < L)$ *be a multi-in-degree vertex, and there exists no multi-in-degree vertex (except* $D$*) above stage* $l$ *(see Figure 6(a)). After applying the* $PA$ *on* $G$*, if* $[ES1_{sub}]^{D}_{S}[L{:}L] = ES2$*, then* $(\forall \langle w, D, L\rangle \in [ES1_{sub}]^{D}_{S})(\exists\ \sigma - path\ S - \cdots - w - D \subseteq ES1_{sub})$*. (the major hard stone; proof see Appendix A.6)*

***The αβ lemma (Summarizing Lemma 1,2).*** *Let* $G =< V, E, S, D, L, \lambda >$ *be the input to the* $PA$*. After applying the* $PA$ *on* $G$*, if* $[ES1_{sub}]^{D}_{S}[L{:}L] = ES2$*, then* $(\forall \langle w, D, L\rangle \in [ES1_{sub}]^{D}_{S})(\exists\ \sigma - path\ S - \cdots - w - D \subseteq ES1_{sub})$.

***Theorem 6 (The sufficiency).*** *If the compact kernel of* $G$ *is not empty, there exists a* $\sigma$*-path in* $G$*. (proof see Appendix A.7, using the* $\alpha\beta$ *lemma)*

Combining Theorem 1,4,5,6, we can eventually prove Conjecture 1.

***Theorem 7 (NP = P).*** *There exists a polynomial-time algorithm for* $2 - MSP$*, i.e., there exists a polynomial-time algorithm for* $NP$*-complete problems.*

## 4 CONCLUDING REMARKS

Simple and mild improvements to known algorithms (Woeginger, 2003; Fomin, & Kaski, 2013) seem just insufficient to break the large complexity barrier. Therefore, we resort to developing our own techniques from scratch—namely, (1) the MSP problem and the proving framework of mathematical induction on the metric $f(G)$ and (2) the ZH algorithm. This is somewhat similar to the research story of the AKS algorithm (Agrawal, Kayal, & Saxena, 2004) for Primality Test. It had been quite shocking on the originality and simplicity of the AKS prime test, given that previous researchers had made much more complicated and modern efforts on theories and methods to attack the problem (often involving great ingenuity); the success was supposed to be contributed to the clever and original combination of classical ideas (Granville, 2004).

The insights of our approach are summarized as follows.

### 4.1 Insights on the MSP problem structure

Unlike other well-known NP-complete problems, the MSP problem is a carefully crafted "unnatural" problem. It is a common practice to concentrate a study on a more convenient novel problem than the original well-known ones— for example, the quasi-polynomial-time lower bound for Graph Isomorphism was obtained when directly solving another polynomial-time equivalent problem (under Karp reductions), i.e., String Isomorphism. The major advantages brought by the structural characteristics of the MSP problem structure is two-fold, as follows.

#### 4.1.1 *The linear-order metric $f(G)$ and the inductive proving framework*

We have been engaged in researching the MSP problem for an extended period, because we have been driven by a fascination with one of its structural properties of MSP. It is believed to be the key towards the design of efficient exact algorithms for the problem.

All MSP instances can be arranged in a sequence according to the quantitative linear-order metric $f(G) = \sum_{v\in V-\{S,D\}}(d^{-}(v)-1)$ (see Metric 1). The problem structure of MSP facilitate the construction of mathematical and algorithmic equivalent instances in the above linear-order sequence, for the inductive proof of the correctness of the algorithm.

Given an arbitrary instance $I_{cur}$ in the sequence. Suppose $d^{-}(v)>1$ for some $v\notin\{S,D\}$ in $I_{cur}$, as shown in Figure 6(a). We can construct an instance $I_{pre}$, such that $I_{cur}$ and $I_{pre}$ keep some sense of mathematical equivalence on the target property concerned by us. The convenience of such a construction originates from the problem structure of MSP: what the construction needs to do, is just following the structure and labels of $I_{cur}$ and defining a different but essentially equivalent set of labels for $I_{pre}$.

This makes it become our persistence to find an algorithm that can fulfill the above proving framework of mathematical induction based on mathematical equivalence—until the ZH algorithm emerged as a solution.

#### 4.1.2 *The system invariant $ES1_{sub}$ and the conservative expansion*

A crucial discovery pertains to a system invariant (i.e., the $ES1_{sub}$ in the PA) between mathematical equivalent MSP instances. This system invariant is used in combination with a "*conservative expansion*" technique, which will be described as follows.

During the inductive proof of the correctness of the proposed ZH algorithm, the algorithm itself is actually used as a "reasoning system". Hence, our primary task is to ensure that the computed results (indeed they are sets of edges) by the actions of ZH algorithm on MSP instances of different order (i.e., the $I_{cur}$ and $I_{pre}$ measured by $f(G)$) can keep essentially the same.

To provide such guarantee for the "reasoning" of the ZH algorithm, we firstly radically expand the labels of $I_{pre}$ (as shown in Figure 6(b)). That is, the labels are expanded to include as many edges as feasible. In this way, verifying the computational results of the ZH algorithm on $I_{pre}$ becomes significantly more straightforward. That's because, according to the reachability of an edge $e$ (i.e., the $R(e)$ defined in the paper), "larger" labels can give $e$ more chances to "go through" the paths in $R(e)$. This parallels the anecdotes of *Isaac Newton's Door with Two Cat Holes*—the little kittens could definitely follow their mother through the larger hole, as long as they can pass through the smaller one. We can hence easily infer the existence of $\sigma$-paths (potential solutions) in $I_{pre}$, by the proposed framework of mathematical induction on $f(G)$.

While the radical expansion provides such convenience, it might potentially bring in extra solutions for $I_{pre}$ when compared with $I_{cur}$ and hence make the two instances become less equivalent. Thus, a control mechanism is requisite to ensure that no more solutions which we care about can be introduced, hence making the radical expansion actually become conservative. The aforementioned system invariant $ES1_{sub}$ serves for this purpose.

The existence of $ES1_{sub}$ has a similar logical power to the existence of uncomputable functions (Church, 1936; Turing, 1936): (1) initially, we use the logical power endowed by the inductive hypothesis to strictly "squeeze out" each such above potential $\sigma$-paths—just an analogy of a function "$f_\alpha(x)$" computed by a Turning machine (represented by the string $\alpha$ and with the input $x$); (2) then, we precisely list out the $\sigma$-paths one by one (as shown in Figure 6(c))—just an analogy of the sequence of all computable functions; (3) finally, we find the system invariant $ES1_{sub}$ for $I_{pre}$ guided by the $ES1_{sub}$ for $I_{cur}$, and further determine the solutions actually demanded by the algorithm through logic inference—just an analogy of the inference of the uncomputable function "$f_x(x) + 1$"; (4) subsequently, the existence of global solutions in $I_{cur}$ can be henceforth constructed.

### 4.2 Insights on the tackling of the complexity

To tackle the hardness, Lance Fortnow (Fortnow, 2009; Fortnow, 2021) categorized some of the tools one can use on NP-complete problems, i.e., brute force (Applegate, Bixby, Chvátal, & Cook, 1998), parameterized complexity (Downey, & Fellows, 2012), approximation (Arora, 1998; Goemans, & Williamson, 1995) and heuristics & average-case complexity (Levin, 1986; The International SAT Competitions, n.d.). Most exact algorithms for NP-complete problems (similar for NP-hard problems) in the literature involve either dynamic programming across the subsets, pruning the search tree, preprocessing the data, or local search (Woeginger, 2003; Fomin, & Kaski, 2013). Though significant progress (including but not limited to Björklund (2014), Björklund, Husfeldt, and Koivisto (2009), Williams (2005)) in the area of exhaustive search has been made in recent decades, existing methods still failed on the formidable exponential barrier. The incapability of those methods mainly lies in that: once losing the help of exhaustive enumeration, the methods just failed to continue to accurately identify the information needed to make the correct global decision. The key to our overcoming of this barrier is two-fold, as follows.

#### *4.2.1 The edge-set representation of paths*

When dealing with paths, traditional exact graph algorithms usually need to explicitly represent each of them as an independent path. Instead, our method treat paths from an edge-set viewpoint, i.e., they are represented by a set of edges traversed by them. Thus, the cost is reduced to polynomial time.

Nevertheless, the representation of paths based on edge sets inevitably introduces ambiguity—a non-empty edge set can be determined by a path, yet the reverse may not hold. An algorithm designed to satisfy our proposed proving framework of mathematical induction on $f(G)$, as described above, provides us a chance to logically prove that: a computed non-empty set of edges by a series of strong constraints (e.g., the compact kernel in the ZH algorithm) can determine the existence of a path with global property (e.g., the demanded $\sigma$-path).

#### *4.2.2 The computation of the reachable-path edge-set $R(e)$, and the discovery of the relation between local and global strategies*

A novel mechanism of the interplay between local strategies and global strategies is discovered and established.

A computational property named the reachability of an edge (i.e., $R(e)$, see Operator 2) is defined and adopted, which can be utilized to summarize the "history" and to detect the "future" for searching "global paths" (i.e., $\sigma$-paths). Contemporarily, the reachability of one edge is constrained by the reachability of the other ones (see Operator 4 and the ZH algorithm). This rightly establishes a recursive relation of the reachability of edges of different stages in the multi-stage graph.

The recursive relation we leverage resembles the state-transition equation in dynamic programming—a standard approach for getting fast exact algorithms for NP-complete problems (Woeginger, 2003; Held, & Karp, 1985; Horowitz, & Sahni, 1978; Lawler, 1976; Lawler, Lenstra, Rinnooy Kan, & Shmoys, 1985; Eppstein, 2001), while the former one appears to be much more convoluted. Nevertheless, since all computations involved can decrease monotonically, such type of algorithm is destined to be polynomial-time upper-bounded.

The proof based on our proposed proving framework of mathematical induction on $f(G)$ provides a robust guarantee for the established recursive relation. The design of the basic operators and the adjustment of $2-$MSP from MSP are also largely driven by logical reasoning to support the proposed proving framework rather than through mere intuition. This is similar to the studies of Ramanujan Summation of "$1+2+3+4+\cdots=-\frac{1}{12}(\Re)$" (Ramanujan, 1903 – 1914), Gödel Incompleteness Theorem (Gödel, 1931), and uncomputable functions (Church, 1936; Turing, 1936), etc., where the motivations and insights were characterized by logical reasoning instead of misleading experiential intuition.

As an aside, it is worth noting that we have tried to rewrite the proofs of several long-existing algorithms using our proposed inductive proving framework. Though we did not discover any brand-new algorithm of better performance, the proving framework did help to find and prove algorithms. For instance, in the case of the Single-Source Shortest Path (SSSP) problem for multi-stage graphs, the correctness of the classic dynamic programming algorithm can be quickly and fluidly verified by mathematical induction on the linear-order metric $f(G)$.

**ACKNOWLEDGMENTS**

This work was partially supported by the National Natural Science Foundation of China (Project: "Research on the Complexity to Solve an NPC Problem", Grant No. 61272010).

**REFERENCES**

Applegate, D., Bixby, R., Chvátal, V., & Cook, W. (1998). On the solution of traveling salesman problems. Documenta Mathematica, Extra Volume ICM III, 645 – 656.

Agrawal, M., Kayal, N., & Saxena, N. (2004). PRIMES is in P. Annals of Mathematics, 160(2), 781 – 793.

Arora, S. (1998). Polynomial time approximation schemes for Euclidean traveling salesman and other geometric problems. Journal of the ACM, 45(5),

753 – 782.

Babai, L. (2016). Graph isomorphism in quasipolynomial time. In Proceedings of the 48th ACM Symposium on the Theory of Computing (pp. 684 – 697).

Björklund, A., Husfeldt, T., & Koivisto, M. (2009). Set partitioning via inclusion-exclusion. SIAM Journal on Computing, 39(2), 546 – 563.

Bürgisser, P., & Ikenmeyer, C. (2011). Geometric complexity theory and tensor rank. In Proceedings of the 43rd ACM Symposium on the Theory of Computing (pp. 509 – 518).

Björklund, A. (2014). Determinant sums for undirected hamiltonicity. In Proceedings of the 51st IEEE Annual Symposium on Foundations of Computer Science (pp. 173 – 182).

Cantor, G. (1874). Über eine Eigenschaft des Inbegriffes aller reellen algebraischen Zahlen. Crelle's Journal, 77, 258 – 262.

Church, A. (1936). A note on the Entscheidungs problem. Journal of Symbolic Logic, 1(1), 40 – 41.

Cook, S. A. (2003). The importance of the P versus NP question. Journal of the ACM, 50(1), 27 – 29.

Downey, R. G., & Fellows, M. R. (2012). Parameterized complexity. Springer Science & Business Media.

Eppstein, D. (2001). Small maximal independent sets and faster exact graph coloring. In Proceedings of the 7th International Workshop on Algorithms and Data Structures (pp. 462 – 470).

Fan, S., Jiang, X., & Peng, L. (2014). Polynomial-time heuristical algorithms for several NP-complete optimization problems. Journal of Computational Information Systems, 10(22), 9707 – 9721.

Fomin, F. V., & Kaski, P. (2013). Exact exponential algorithms. Communications of the ACM, 56(3), 80 – 88.

Fortnow, L. (2009). The status of the P versus NP problem. Communications of the ACM, 52(9), 78 – 86.

Fortnow, L. (2021). Fifty years of P vs. NP and the possibility of the impossible. Communications of the ACM, 65(1), 76 – 85.

Furst, M., Saxe, J. B., & Sipser, M. (1984). Parity, circuits and the polynomial-time hierarchy. Mathematical Systems Theory, 17, 13 – 27.

Garey, M. R., & Johnson, D. S. (1979). Computers and intractability: A guide to the theory of NP-completeness. Freeman.

Gödel, K. (1931). Über formal unentscheidbare Sätze der Principia Mathematica und verwandter Systeme I. Monatshefte für Mathematik Physik, 38, 173 – 198.

Granville, A. (2004). It is easy to determine whether a given integer is prime. Bulletin (New Series) of the American Mathematical Society, 42(1), 3 – 38.

Goemans, M. X., & Williamson, D. P. (1995). Improved approximation algorithms for maximum cut and satisfiability problems using semidefinite programming. Journal of the ACM, 42(6), 1115 – 1145.

Haken, A. (1985). The intractability of resolution. Theoretical Computer Science, 39, 297 – 305.

Held, M., & Karp, R. M. (1985). A dynamic programming approach to sequencing problems. Journal of the Society for Industrial and Applied Mathematics, 10(1), 196 – 210.

Horowitz, E., & Sahni, S. (1978). Fundamentals of computer algorithms. Computer Science Press.

Jiang, X. (2020). Polynomial-time algorithm for Hamilton Circuit problem. Computer Science, 47(7), 8 – 20. (In Chinese with English abstract)

Jiang, X., Liu, W., Wu, T., & Zhou, L. (2014). Reductions from MSP to SAT and from SUBSET SUM to MSP. Journal of Computational Information Systems, 10(3), 1287 – 1295.

Jiang, X., Peng, L., & Wang, Q. (2010). MSP problem: Its NP-completeness and its algorithm. In Proceedings of the 5th IEEE International Conference on Ubiquitous Information Technologies and Applications (pp. 1 – 5).

Karp, R. M. (1972). Reducibility among combinatorial problems. In Complexity of Computer Computations (pp. 85 – 103).

Knuth, D. (2002). All questions answered. Notices of the AMS, 49(3), 318 – 324.

Lawler, E. L. (1976). A note on the complexity of the chromatic number problem. Information Processing Letters, 5(3), 66 – 67.

Levin, L. A. (1986). Average case complete problems. SIAM Journal on Computing, 15, 285 – 286.

Lawler, E. L., Lenstra, J. K., Rinnooy Kan, A. H. G., & Shmoys, D. B. (1985). The traveling salesman problem: A guided tour of combinatorial optimization. Wiley-Interscience.

van Melkebeek, D. (2007). A survey of lower bounds for satisfiability and related problems. Foundations and Trends in Theoretical Computer Science,

197 – 303.

Mulmuley, K. D., & Sohoni, M. (2001). Geometric complexity theory I: An approach to the P vs. NP and related problems. SIAM Journal on Computing, 31(2), 496 – 526.

Mulmuley, K. D. (2012). The GCT program toward the P vs. NP problem. Communications of the ACM, 55(6), 98 – 107.

Ramanujan, S. (1903 – 1914). Second notebook (Unpublished, Chapter VI).

Razborov, A. A. (1985). Lower bounds on the monotone complexity of some boolean functions. Soviet Mathematics – Doklady, 31, 485 – 493.

Razborov, A. A. (1989). On the method of approximations. In Proceedings of the 21st ACM Symposium on the Theory of Computing (pp. 167 – 176).

Razborov, A. A., & Rudich, S. (1997). Natural proofs. Journal of Computer and System Sciences, 55(1), 24 – 35.

The international SAT competitions. (n.d.). Retrieved February 11, 2023, from http://www.satcompetition.org

Baker, T., Gill, J., & Solovay, R. (1975). Relativizations of the P = NP question. SIAM Journal on Computing, 4(4), 431 – 442.

Turing, A. M. (1936). On computable numbers, with an application to the Entscheidungs problem. In Proceedings of the London Mathematical Society (Vol. 42, pp. 230 – 265).

Valiant, L. G. (2002). Quantum circuits that can be simulated classically in polynomial time. SIAM Journal on Computing, 31(4), 1229 – 1254.

Viola, E. (2018, February 16). I believe P = NP. Retrieved September 29, 2022, from https://emanueleviola.wordpress.com/2018/02/16/i-believe-pnp/

Williams, R. (2005). A new algorithm for optimal 2-constraint satisfaction and its implications. Theoretical Computer Science, 348(2 – 3), 357 – 365.

Woeginger, G. J. (2003). Exact algorithms for NP-hard problems: A survey. In Combinatorial Optimization—Eureka, You Shrink! (Lecture Notes in Computer Science, Vol. 2570, pp. 185 – 207).

Wöginger, G. J. (n.d.). The P-versus-NP page. Retrieved January 7, 2022, from https://www.win.tue.nl/~gwoegi/P-versus-NP.htm

Xu, K., Boussemart, F., Hemery, F., & Lecoutre, C. (2007). Random constraint satisfaction: Easy generation of hard (satisfiable) instances. Artificial Intelligence, 171(8 – 9), 514 – 534.

Xu, K., & Li, W. (2000). Exact phase transitions in random constraint satisfaction problems. Journal of Artificial Intelligence Research, 12(1), 93 – 103.

Garey, M. R., & Johnson, D. S. (1979). Computers and intractability: A guide to the theory of NP-completeness. W. H. Freeman.

## A APPENDICES

### A.1 Proof of Theorem 1

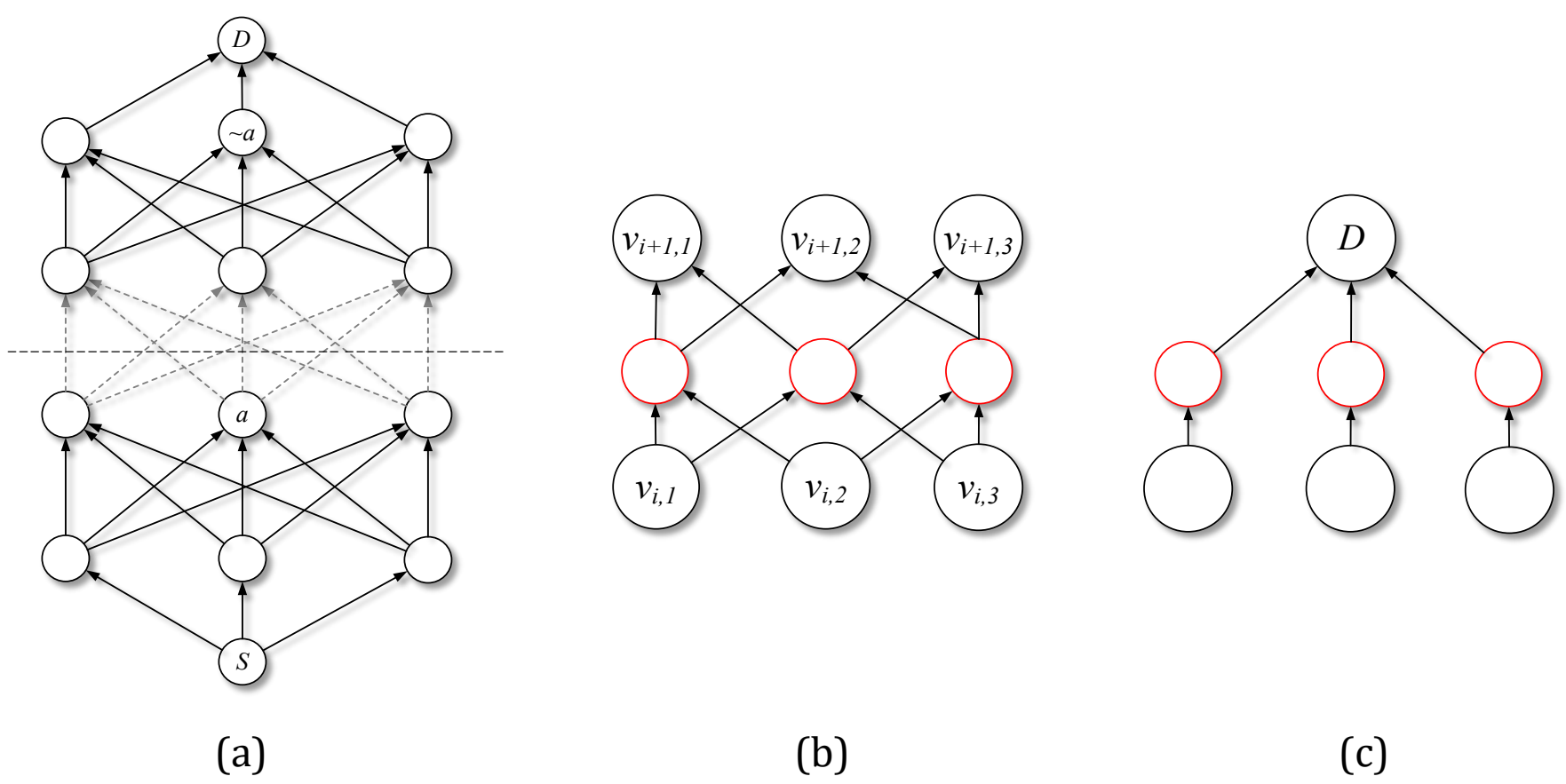

(a) (b) (c)

Figure 7: Reduction from 3-SAT to 2-MSP

*Proof.* A number of NP-complete problems can be polynomially reduced to the MSP problem (see Jiang, Liu, Wu, and Zhou (2014), Fan, Jiang, and Peng (2014)). Jiang, Liu, Wu, and Zhou (2014) presented the following reduction from $\mathrm{CNF}-\mathrm{SAT}$:

(i) Generate a vertex of $G = \langle V, E, S, D, L, \lambda \rangle$ in a MSP instance, for each literal of a clause in a $\mathrm{CNF}-\mathrm{SAT}$ instance.

(ii) Relate each clause in the $\mathrm{CNF}-\mathrm{SAT}$ instance to one stage of vertices in $G$.

(iii) Add two vertices $S$ and $D$ to $G$.

(iv) Add edges to make all vertices between adjacent stages fully connected.

(v) Set $\lambda(D) = E$; for each vertex $x \in V - \{S, D\}$ (assume $x$ corresponds to some literal $p$), set $\lambda(x) = E - \left\{ e \,\middle|\, \begin{array}{l} e \text{ starts from } \bar{x} \text{ or end at } \bar{x}, \text{where the vertex } \bar{x} \\ \text{corresponds to the complementary literal of } p \end{array} \right\}$.

An example of the reduction from $3-\mathrm{SAT}$ to MSP is illustrated in Figure 7(a). In this sense, MSP establishes a graph model for $\mathrm{CNF}-\mathrm{SAT}$.

To further reduce $3-\mathrm{SAT}$ to $2-\mathrm{MSP}$, we just need to replace edges between each pair of adjacent stages (except the first two and last two stages) with a stage gadget as shown in Figure 7(b). The three vertices at the lower stage (e.g., $v_{i,1}, v_{i,2}, v_{i,3}$ in Figure 7(b)) are organized via the combination of "$C_3^2$" to "connect to" the three auxiliary vertices in the gadget; and then the auxiliary vertices "connect to" the vertices at the upper stage (e.g., $v_{i+1,1}, v_{i+1,2}, v_{i+1,3}$ in Figure 7(b)) in the same mode. Assume the auxiliary vertices are $a_1, a_2, a_3$ and now belong to stage $l$, then the new edges are $\langle v_{i,1}, a_1, l \rangle$, $\langle v_{i,1}, a_2, l \rangle$, $\langle v_{i,2}, a_1, l \rangle$, $\langle v_{i,2}, a_3, l \rangle$, $\langle v_{i,3}, a_2, l \rangle$, $\langle v_{i,3}, a_3, l \rangle$, $\langle a_1, v_{i+1,1}, l+1 \rangle$, $\langle a_1, v_{i+1,2}, l+1 \rangle$, $\langle a_2, v_{i+1,1}, l+1 \rangle$, $\langle a_2, v_{i+1,3}, l+1 \rangle$, $\langle a_3, v_{i+1,2}, l+1 \rangle$, $\langle a_3, v_{i+1,3}, l+1 \rangle$.

In addition, replace the edges $\langle v_{L-1,1}, D, L \rangle$, $\langle v_{L-1,2}, D, L \rangle$, $\langle v_{L-1,3}, D, L \rangle$ between the last two stages with a stage gadget as shown in Figure 7(c). An auxiliary vertex is inserted between a vertex at the lower stage and the sink $D$.

Assume the auxiliary vertices are $a_1$, $a_2$, $a_3$, then the new edges are $\langle v_{L-1,1}, a_1, L\rangle$, $\langle a_1, D, L+1\rangle$, $\langle v_{L-1,2}, a_2, L\rangle$, $\langle a_2, D, L+1\rangle$, $\langle v_{L-1,3}, a_3, L\rangle$, $\langle a_3, D, L+1\rangle$.

The labels of the auxiliary vertices can all be set to the updated $E$. The labels of the original vertices should be recomputed (by the above step (v) of the reduction from CNF − SAT to MSP) to accommodate changes in $E$.

A complete view of the reduction is illustrated in Figure 1(c).

It takes little effort to exam just item-by-item, that the resulted instance fulfills Definition 3(b). Note that, in the case of 2 − MSP, we can assume each CNF consists of at least 2 clauses.

The 3 − SAT problem is therefore polynomial-time Karp-reducible (Karp, 1972) to 2 − MSP.

For one direction, it's easy to see by Definition 2 that, the 3 − SAT instance can be satisfied, if there exists some σ-path in the corresponding 2 − MSP instance.

In fact, if $v_0 - v_1 - v_2 - \cdots - v_L \subseteq E$ (where $v_0 = S$, $v_L = D$) is a σ-path, then each $v_{2*i-1}$ ($i \in \{1,2,\ldots,N\}$, $N \geq 2$ is the number of clauses) on the path must be a vertex that stands for a literal (let it denoted by $p_{v_{2*i-1}}$) in a clause $C_i$ of the 3 − SAT problem, according to the above reduction.

Since each $\lambda(v_{2*i-1})$ ($i \in \{1,2,\ldots,N\}$) excludes edges that start from or end at $\bar{x}$ (where $\bar{x}$ corresponds to the complementary literal of $p_{v_{2*i-1}}$), no pair of literals among $\left\{p_{v_{2*j-1}} \middle| j \in \{1,2,\ldots,N\}\right\}$ is complementary. We thus know that there must exist an assignment φ that satisfies all these $p_{v_{2*j-1}}$ ($j \in \{1,2,\ldots,N\}$). This assignment φ also satisfies the given 3 − SAT problem.

For the other direction, if the 3 − SAT instance is satisfied by some assignment φ, then there must exist some literal $p_i$ ($i \in \{1,2,\ldots,N\}$, $N \geq 2$ is the number of clauses) in each clause $C_i$ such that $\varphi(p_i) = \text{true}$.

Then, by the above definition of the labels and by Definition 2, the path $S - x(p_1) - x_1 - x(p_2) - x_2 - \cdots - x(p_N) - x_N - D$ (where $x(p_i)$ stands for the vertex created corresponding to $p_i$, $x_i$ stands for an inserted auxiliary vertex, $i \in \{1,2,\ldots,N\}$) in the corresponding 2 − MSP instance must be a σ-path, according to the reduction.

Therefore, 2 − MSP is NP-complete.☐

The verification of the NP-completeness of MSP is so trivial even for non-specialist readers, that we have published more than ten proofs and even assigned it as a small homework to hundreds of graduate students in an algorithms & complexity course for many consecutive years. In several seminars when visiting other universities, tens of students proposed at least 6 independent approaches of reduction, as we know.

### A.2 The theorems of equivalence & uniqueness for the basic operators

We now rewrite Operator 3,4 in "analytic forms".

***Operator 3 ($\chi^{v}_{R(E)}(ES)$), analytic form).*** *Given $ES \subseteq E$, $v \in V_l$ and the collection of ρ-path edge-sets $R(E)$. Give Operator 3 as the "analytic form": $\chi^{v}_{R(E)}(ES) =_{def} ES1$, s.t. $ES1 \subseteq ES$, $\mathfrak{X}(ES1) = true$ and $\mathfrak{X}(ES2) = false$ ($ES2 \subseteq ES$) for each possible $ES2 \supset ES1$, where*

$$\mathfrak{X}(\Delta) =_{\text{def}} \left( \Delta = \left[ \left\{ \begin{matrix} e = \langle a,b,k\rangle \\ \in \Delta \end{matrix} \middle| \begin{matrix} [R(e) \cap \Delta]^{v}_{b} \neq \emptyset \text{ (when } k < l\text{);} \\ [R(e)]^{D}_{v} \neq \emptyset \text{ (when } k = l \neq L\text{)} \end{matrix} \right\} \right]^{v}_{S} \right) (\Delta \subseteq E). \quad (4)$$

***Operator 4 ($\boldsymbol{\psi_{R(E)-\{R(e)\}}(R(e))}$, analytic form).*** *Given $e = \langle u, v, l \rangle \in E$ $(1 < l < L)$ and the collection of $\rho$-path edge-sets $R(E)$. Give the operator $\psi_{R(E)-\{R(e)\}}(R(e))$ as the "analytic form": $\psi_{R(E)-\{R(e)\}}(R(e))$ modifies $R(e)$ into an edge set $ES1$, s.t. $ES1 \subseteq R(e)$, $\mathfrak{Y}(ES1) = true$ and $\mathfrak{Y}(ES2) = false$ $(ES2 \subseteq R(e))$ for each possible $ES2 \supset ES1$, where*

$$\mathfrak{Y}(\Delta) =_{\text{def}} \left( \Delta = \left[ \left\{ \begin{matrix} e' = \langle a, b, k \rangle \\ \in \Delta \end{matrix} \middle| \begin{matrix} \boldsymbol{B} \neq \emptyset, \text{where} \\ \boldsymbol{A} = \chi_{R(E)}^{b}\left(\left\{\langle x, y, i \rangle \in E \middle| e' \in [R(\langle x, y, i \rangle) \cap \lambda(b)]_{y}^{b}\right\} \cup \{e'\}\right) \\ \text{and } \boldsymbol{B} = \chi_{R(E)}^{u}\left(\left\{\langle c, d, j \rangle \in \boldsymbol{A} \middle| \{e, e'\} \subseteq [R(\langle c, d, j \rangle) \cap \boldsymbol{A}]_{d}^{b}\right\}\right) \end{matrix} \right\} \right]_{v}^{D} \right) \tag{5}$$

$(\Delta \subseteq E)$.

The following theorems ensure that the two distinct operator definitions coincide. In other words, the intended algorithmic operations underpinning the proposed ZH algorithm are all precisely and uniquely determined.

***Theorem 2 (Equivalence & uniqueness, Operator 3).*** *The "analytic form" and the "procedural form" define the same operator. In other words, there exists only one unique edge set that fulfills the "analytic form" (or the "procedural form") of Operator 3.*

*Proof.* Let $\Delta = [\Delta]_{S}^{v} \subseteq ES$ be an edge set that fulfills its "analytic form", and let $\Delta' = [\Delta']_{S}^{v} \subseteq ES$ be the result of its "procedural form".

(1) Case $\Delta \subset \Delta'$: according to the computation of procedural form, for some $\langle a, b, k \rangle \in (\Delta' - \Delta) \subseteq ES$, we must have $\begin{cases} [R(e) \cap \Delta]_{b}^{v} \neq \emptyset \text{ (when } k < l) \\ [R(e)]_{v}^{D} \neq \emptyset \text{ (when } k = l \neq L) \end{cases}$. This contradicts with the definition of $\Delta$.

(2) Case $\Delta' \subset \Delta$: any $\langle a, b, k \rangle \in (\Delta - \Delta') \subseteq ES$ will be deleted according to the computation of procedural form, since $\begin{cases} [R(e) \cap \Delta]_{b}^{v} = \emptyset \text{ (when } k < l) \\ [R(e)]_{v}^{D} = \emptyset \text{ (when } k = l \neq L) \end{cases}$ violates the analytic form.

(3) Any other case, $\Delta \subset \Delta \cup \Delta'$. This will violate the definition of analytic form.☐

***Theorem 3 (Equivalence & uniqueness, Operator 4).*** *The "analytic form" and the "procedural form" define the same operator. In other words, there exists only one unique edge set that fulfills the "analytic form" (or the "procedural form") of Operator 4.*

*Proof.* Let $\Delta = [\Delta]_{v}^{D} \subseteq R(e)$ be an edge set that fulfills its "analytic form", and let $\Delta' = [\Delta']_{v}^{D} \subseteq R(e)$ be the result of its "procedural form".

(1) Case $\Delta \subset \Delta'$: according to the computation of procedural form, for some $\langle a, b, k \rangle \in (\Delta' - \Delta) \subseteq R(e)$, we must have $\boldsymbol{A} = \chi_{R(E)}^{b}\left(\left\{\langle x, y, i \rangle \in E \middle| e' \in [R(\langle x, y, i \rangle) \cap \lambda(b)]_{y}^{b}\right\} \cup \{e'\}\right) \neq \emptyset$ and $\boldsymbol{B} = \chi_{R(E)}^{u}\left(\left\{\langle c, d, j \rangle \in \boldsymbol{A} \middle| \{e, e'\} \subseteq [R(\langle c, d, j \rangle) \cap \boldsymbol{A}]_{d}^{b}\right\}\right) \neq \emptyset$. This contradicts with the definition of $\Delta$.

(2) Case $\Delta' \subset \Delta$: any $\langle a, b, k \rangle \in (\Delta - \Delta') \subseteq R(e)$, will be deleted according to the computation of procedural form, since $\boldsymbol{A} = \chi_{R(E)}^{b}\left(\left\{\langle x, y, i \rangle \in E \middle| e' \in [R(\langle x, y, i \rangle) \cap \lambda(b)]_{y}^{b}\right\} \cup \{e'\}\right) = \emptyset$ and $\boldsymbol{B} = \chi_{R(E)}^{u}\left(\left\{\langle c, d, j \rangle \in \boldsymbol{A} \middle| \{e, e'\} \subseteq [R(\langle c, d, j \rangle) \cap \boldsymbol{A}]_{d}^{b}\right\}\right) = \emptyset$.

(3) Any other case, $\Delta \subset \Delta \cup \Delta'$. This will violate the definition of analytic form.☐

#### A.3 Proof of Theorem 4

*Proof.* Each size of $\lambda(v)$ and $R(e)$ is no more than $|E|$. Moreover, $|R(E)|$ is no more than $|E|$.

The cost for computing $\chi^{v}_{R(E)}(ES)$ can be $O(|E|^4)$ and the cost for computing $\psi_{R(E)-\{R(e)\}}\big(R(e)\big)$ can be $O(|E|^7)$, hence the cost of step 2 of the ZH *a*lgorithm can be $O(|E|^8)$.

Step 2 is the most expensive statement in the ZH algorithm. Each iteration of step 2 will prune at least one edge in $R(\langle u,v,l\rangle)$, and the number of edges each in $R(\langle u,v,l\rangle)$ and $\lambda(v)$ is no more than $|E|$. The number of $R(e)$ is $|R(E)|$. So, the cost of step 2 and step 3 can be $|E| * |R(E)| * O(|E|^8)$.

Overall, the cost of the ZH algorithm can be $O(|E|^{10})$, a polynomial function of $|E|$.☐

#### A.4 Proof of Theorem 5

*Proof.* Let $P = v_0 - v_1 - v_2 - \cdots - v_L$ be a $\sigma$-path in $G$, where $v_0 = S$ and $v_L = D$. By Definition 2, $[P]^{v_h}_{v_0} \subseteq \lambda(v_h)$ $(1 \le h \le L)$, and for $\langle v_{l-1}, v_l, l\rangle \in P$ $(1 \le l \le L)$ we have $\langle v_{l-1}, v_l, l\rangle \in \lambda(v_l) \cap \lambda(v_{l+1}) \cap \cdots \cap \lambda(D)$. Thus, after the execution of step 1 of the ZH algorithm, we have $[P]^{v_L}_{v_l} \subseteq R(\langle v_{l-1}, v_l, l\rangle)$ $(1 \le l \le L)$. After step 2, we still have $[P]^{v_L}_{v_l} \subseteq R(\langle v_{l-1}, v_l, l\rangle)$ $(1 \le l \le L)$. Step 3 can not prune any path in $R(\langle v_{l-1}, v_l, l\rangle)$. This ensure that $P \subseteq \chi^{D}_{R(E)}(\lambda(D))$. Hence, $\chi^{D}_{R(E)}(\lambda(D)) \ne \emptyset$.☐

#### A.5 Proof of Lemma 1

*Proof.* $[ES1_{sub}]^{D}_{S}[L\!:L] = ES2 \ne \emptyset$ implies $ES1 = \chi^{D}_{R(E)}(ES1) \ne \emptyset$. Then, for $\langle a_{L-1}, D, L\rangle \in ES2$, by the definition of Operator 3, there exists $\langle a_{L-2}, a_{L-1}, L-1\rangle \in ES1$ such that $\langle a_{L-1}, D, L\rangle \in R(\langle a_{L-2}, a_{L-1}, L-1\rangle) \cap \chi^{D}_{R(E)}(ES1)$. According to the computation of $\psi_{R(E)-\{R(\langle a_{L-2},a_{L-1},L-1\rangle)\}}(R(\langle a_{L-2}, a_{L-1}, L-1\rangle))$, when determining "$\langle a_{L-1}, D, L\rangle \in R(\langle a_{L-2}, a_{L-1}, L-1\rangle)$", we have

$$\boldsymbol{A} = \chi^{D}_{R(E)}\left(\begin{array}{c}\left\{\langle x,y,i\rangle \in E \,\middle|\, \begin{array}{c}\langle a_{L-1}, D, L\rangle \in \\ [R(\langle x,y,i\rangle) \cap \lambda(D)]^{D}_{y}\end{array}\right\} \\ \cup \{\langle a_{L-1}, D, L\rangle\}\end{array}\right) \ne \emptyset. \tag{6}$$

By the definition of Operator 3 and by the fact that no multi-in-degree vertex can be found in $G$ from stage 1 to stage $L-1$, only one single preceding edge of stage $l-1$ can be found for each edge of stage $l$ $(2 \le l \le L)$ in $\boldsymbol{A}$. Thus, the set $\boldsymbol{A}$ is uniquely determined as the path $S - \cdots - a_{L-2} - a_{L-1} - D$, which must be a $\sigma$-path and $S - \cdots - a_{L-2} = [ES1]^{a_{L-2}}_{S}$.

The above discussion based on Operator 3 and on the graph structure also implies that, (i) each path $S - \cdots - D \subseteq ES1$ must be a $\sigma$-path and (ii) each $S - \cdots - D \subseteq [ES1_{sub}]^{D}_{S} \subseteq ES1$ must also be a $\sigma$-path.☐

#### A.6 Proof of Lemma 2

To prove Lemma 2, we need to construct a graph $G' =< V', E', S, D, L, \lambda' >$, such that: (1) $f(G') < f(G)$, and $G'$ satisfies Definition 3(b); (2) if ($[ES1_{sub}]^{D}_{S}[L\!:L]$ of $G$) = ($ES2$ of $G$) for $G$, then ($[ES1_{sub}]^{D}_{S}[L\!:L]$ of $G'$) = ($ES2$ of $G'$) for $G'$; (3) if ($SP$ of $G'$) $\subseteq$ ($ES1_{sub}$ of $G'$) is a solution required by the PA for $G'$, then some solution ($SP$ of $G$) $\subseteq$ ($ES1_{sub}$ of $G$) required by the PA for $G$ must exist.

### *A.6.1 The construction of a less equivalent $G_1$ and the proof of Claim 1,2,3,4*

For the multi-in-degree vertex $v \in V_l$ $(1 < l < L)$ specified by Lemma 2, we have $d^+(v) > 0$ and $d^-(v) = 2$ by Definition 3(b). Moreover, $d^-(t) = \cdots = d^-(w) = 1$ for each path from $v$ to $D$ like $v - t - \cdots - w - D$ (the path can be shorter than 3 edges, as the introduction of the additional vertex "$t$" just helps the illustration but is not a must). Assume $\langle u_1, v, l\rangle$ and $\langle u_2, v, l\rangle$ are just the two edges ending at $v$, as shown in Figure 6(a).

By Definition 3(b) (item 2), we have $2 \le l \le L - 2$ (recall that $L \ge 5$ by Definition 1).

Based on $G$, we can construct a new graph $G_1 =< \widehat{V_1}, E_1, S, D, L, \lambda_1 >$ as follows, by "splitting" the multi-in-degree vertex $v$. It should be noted that, when defining a graph $G_n$ $(n \in \mathbb{N})$, we use $\widehat{V_n}$ to represent the set of all vertices in $G_n$, instead of $V_n$. In this way, $\widehat{V_n}$ can be distinguished from the "$V_n$" in Definition 1 (recall that $V_n$ represents the set of all vertices at a specified stage $n \in \{0, \ldots, L\}$).

To define $\widehat{V_1}$ and $E_1$, we delete all paths $u_i - v - \cdots - D$ $(i = 1,2)$ in $G$ and add two new paths $u_i - v_i - t_i - \cdots - r_i - w_i - D$ $(i = 1,2)$. Keep all the rest vertices and edges unchanged. We then get the structure of an $L$ −stage graph $G_1$, as shown in Figure 6(b).

To define $\lambda_1$ (i.e., $(\lambda(x)$ of $G_1)$ for $x \in \widehat{V_1}$):

(i) For $x = D$, let $(\lambda(D)$ of $G_1) = E_1$.

(ii) For $x \in V - \{v_1, v_2, t_1, t_2, \ldots, w_1, w_2, D\}$, let $(\lambda(x)$ of $G_1) = (\lambda(x)$ of $G) \cap [E]_S^x$.

(iii) For $x \in \{v_1, v_2, t_1, t_2, \ldots, w_1, w_2\}$, the following **radical expansion** of labels is done:

- For $x \in \{v_1, v_2\}$, set $(\lambda(v_i)$ of $G_1) = \{\langle u_i, v_i, l\rangle\} \cup \begin{pmatrix} (\lambda(v) \text{ of } G)[1{:}\, l-1] \\ \cap\, (\lambda(u_i) \text{ of } G)[1{:}\, l-1] \end{pmatrix}$ $(i = 1,2)$.
- For $x$ on $t_1 - \cdots - w_1$ and $t_2 - \cdots - w_2$, set $(\lambda(t_i)$ of $G_1) = (\lambda(v_i)$ of $G_1) \cup \{\langle v_i, t_i, l+1\rangle\}$, ..., $(\lambda(w_i)$ of $G_1) = (\lambda(v_i)$ of $G_1) \cup (v_i - t_i - \cdots - r_i - w_i)$ $(i = 1,2)$.

The above "split" of $v$ does not damage the existence of the original $\sigma$-paths. Indeed, the radical expansion makes it easy to confirm the computation and the result of $R(e)$ after applying $\psi_{R(E)-\{R(e)\}}(R(e))$, when $G_1$ is the input to the PA.

However, since $(\lambda(x)$ of $G_1)$ $(x \in \widehat{V_1})$ seems to contain more edges in essence than its counterpart in $G$, if $P$ is a $\sigma$-path in $G_1$, maybe no $\sigma$-path corresponding to $P$ exists in $G$. Nevertheless, in the current situation, this won't cause troubles (see the following Claim 1,2,3,4). Moreover, it will be proved that, if there is a $\sigma$-path as claimed by step 3 of the PA for the "smaller" graph, then there must exist a $\sigma$-path as claimed by step 3 of the PA for $G$ (see the following Claim 5). Hence, the constraints posed by $(ES1_{sub}$ of $G_1)$ (if non-empty) manage to suppress the undesired solutions introduced by the radical expansion (and the expansion now actually becomes **conservative**).

Subsequently, the construction of $G_1$ is completed.

Now we divide Lemma 2 into the following Claim 1,2,3,4,5. Implicitly, these claims share the same context of Lemma 2.

***Claim 1 (for $G_1$).*** $f(G_1) < f(G)$.

*Proof.* Since $v$ is the multi-in-degree vertex that appears at stage $l$ $(1 < l < L)$ and no multi-in-degree vertex (except $D$) can be found above stage $l$, we have

$$
\begin{aligned}
&\sum\nolimits_{x \in (V_l \text{ of } G_1)} (d^-(x) - 1) \\
&= \sum\nolimits_{x \in (V_l \text{ of } G_1) - \{v_1, v_2\}} (d^-(x) - 1) + (d^-(v_1) - 1) + (d^-(v_2) - 1) \\
&\leq \sum\nolimits_{x \in (V_l \text{ of } G) - \{v\}} (d^-(x) - 1) + (d^-(v) - 1) - 1 \\
&= \sum\nolimits_{x \in (V_l \text{ of } G)} (d^-(x) - 1) - 1\,.
\end{aligned}
\tag{7}
$$

Therefore, $f(G_1) < f(G)$.☐

***Claim 2 (for $G_1$).*** *If $\left(\chi_{R(E)}^{D}(ES1) \text{ of } G\right) \neq \emptyset$, we have*

$$
(ES1 \text{ of } G_1) \supseteq \begin{pmatrix} \left((ES1 \text{ of } G) - \left\{e \middle| e \in u_i - v - \cdots\cdots - D \subseteq E, i \in \{1,2\}\right\}\right) \\ \cup \left\{e \middle| \begin{matrix} e \in u_i - v_i - \cdots - w_i - D \subseteq E_1, \\ (R(\langle u_i, v, l \rangle) \text{ of } G) \neq \emptyset, i \in \{1,2\} \end{matrix} \right\} \end{pmatrix} \neq \emptyset. \tag{8}
$$

*Proof.* For every $\langle r, s, k \rangle$ and $\langle o, p, h \rangle$ $(1 \leq k < h \leq L)$ in $G$: (i) if the initial $\rho$-path edge-set $(R_0(\langle r, s, k \rangle) \text{ of } G)$ contains $\langle o, p, h \rangle$, there must exist some $e_1$ and $e_2$ in $G_1$, such that $e_2 \in (R_0(e_1) \text{ of } G_1)$; (ii) if the $(R(\langle r, s, k \rangle) \text{ of } G) \in (R(E) \text{ of } G)$ computed by the ZH algorithm contains $\langle o, p, h \rangle$, according to the radical expansion of $G_1$, there must exist $e_1$ and $e_2$ in $G_1$, such that the $(R(e_1) \text{ of } G_1) \in (R(E) \text{ of } G_1)$ computed by the ZH algorithm contains $e_2$.[3]

If the above $\langle r, s, k \rangle$ and $\langle o, p, h \rangle$ are associated with the vertices involved in $[E]_v^D[l+1{:}L]$, then $e_1$ and $e_2$ are associated with the vertices (of the corresponding stages) involved in $[E_1]_{v_1}^D[l+1{:}L] \cup [E_1]_{v_2}^D[l+1{:}L]$, otherwise $e_1 = \langle r, s, k \rangle$ and $e_2 = \langle o, p, h \rangle$. For instance, when $\langle o, p, h \rangle \in v - \cdots\cdots - D \subseteq E$ and $k < l$, $\langle o, p, h \rangle \in (R(\langle r, s, k \rangle) \text{ of } G)$ implies $\langle u_i, v, l \rangle \in (R(\langle r, s, k \rangle) \text{ of } G)$ (for some $i \in \{1,2\}$); then by the radical expansion, we can have $\langle u_i, v_i, l \rangle \in (R(\langle r, s, k \rangle) \text{ of } G_1)$ and further we can deduce that there exists an edge $e_2 \in v_i - \cdots - w_i - D \subseteq E_1$ of stage $h$ in $G_1$ such that $e_2 \in (R(\langle r, s, k \rangle) \text{ of } G_1)$.

More discussions on $(R(E) \text{ of } G_1)$ and the detailed renaming rules for the above $e_1$ and $e_2$, if needed, are provided in Appendix A.8.1.

With the above clarification of $(R(E) \text{ of } G_1)$, then by the definition of Operator 3, we can hence obtain

$$
\left(\chi_{R(E)}^{D}(\lambda(D)) \text{ of } G_1\right) \supseteq \begin{pmatrix} \left(\chi_{R(E)}^{D}(\lambda(D)) \text{ of } G\right) - \\ \left\{e \middle| e \in u_i - v - \cdots\cdots - D \subseteq E, i \in \{1,2\}\right\} \end{pmatrix}. \tag{9}
$$

Note that, we have $\langle u_i, v, l \rangle \in \left(\chi_{R(E)}^{D}(\lambda(D)) \text{ of } G\right)$ when $(R(\langle u_i, v, l \rangle) \text{ of } G) \neq \emptyset$ ( $i \in \{1,2\}$ ), because $\emptyset \neq \left(\chi_{R(E)}^{D}\begin{pmatrix} \boldsymbol{A}[l+1{:}L] \cup \\ \{\langle u_i, v, l \rangle\} \cup \boldsymbol{B} \end{pmatrix} \text{ of } G\right) \subseteq \left(\chi_{R(E)}^{D}(\lambda(D)) \text{ of } G\right)$ (where $(\boldsymbol{A} \text{ of } G)$, $(\boldsymbol{B} \text{ of } G)$ are the sets computed when deciding to preserve some $\langle w, D, L \rangle \in E$ in $(R(\langle u_i, v, l \rangle) \text{ of } G)$ by Operator 4, see step 2 of $ZH\backslash step4$) by bottom-up checking the edges in the definition of Operator 3 and by leveraging the fact that $(\boldsymbol{A} \text{ of } G) \subseteq (\lambda(D) \text{ of } G)$.

[3] Unless otherwise specified, $R(e)$ refers to the stable one after step 2 of $ZH\backslash step4$, since it is the minimum.

Further note that, $(R(\langle u_i, v, l\rangle) \text{ of } G) \neq \emptyset$ $(i \in \{1,2\})$ implies $\langle w, D, L\rangle \in (R(\langle u_i, v, l\rangle) \text{ of } G)$. That further implies $\langle w_i, D, L\rangle \in (R(\langle u_i, v_i, l\rangle) \text{ of } G_1)$, and hence $v_i - t_i - \cdots - r_i - w_i - D \subseteq (R(\langle u_i, v_i, l\rangle) \text{ of } G_1)$ by the radical expansion.

Subsequently, by the definition of Operator 3, we can have

$$\left(\chi_{R(E)}^{D}\big(\lambda(D)\big) \text{ of } G_1\right) \supseteq \left\{e \,\middle|\, \begin{matrix} e \in u_i - v_i - \cdots - w_i - D \subseteq E_1, \\ (R(\langle u_i, v, l\rangle) \text{ of } G) \neq \emptyset, i \in \{1,2\} \end{matrix} \right\}. \tag{10}$$

Summarizing all above discussions, we now obtain

$$\begin{gathered} \left(\chi_{R(E)}^{D}\big(\lambda(D)\big) \text{ of } G_1\right) \supseteq \\ \left( \begin{gathered} \left(\left(\chi_{R(E)}^{D}\big(\lambda(D)\big) \text{ of } G\right) - \left\{e \,\middle|\, e \in u_i - v - \cdots\cdots - D \subseteq E, i \in \{1,2\}\right\}\right) \\ \cup \left\{e \,\middle|\, \begin{matrix} e \in u_i - v_i - \cdots - w_i - D \subseteq E_1, \\ (R(\langle u_i, v, l\rangle) \text{ of } G) \neq \emptyset, i \in \{1,2\} \end{matrix} \right\} \end{gathered} \right). \end{gathered} \tag{11}$$

As a result, $\left(\chi_{R(E)}^{D}\big(\lambda(D)\big) \text{ of } G\right) \neq \emptyset$ would imply $\left(\chi_{R(E)}^{D}(ES1) \text{ of } G_1\right) = (ES1 \text{ of } G_1) = \left(\chi_{R(E)}^{D}\big(\lambda(D)\big) \text{ of } G_1\right) \neq \emptyset$.☐

***Claim 3 (for $G_1$).*** *$G_1$ satisfies Definition 3(b).*

*Proof.* Recall that $G$ fulfills Definition 3(b).

It is easy to check item-by-item that, the “split” of $v$ won't violate Definition 3(b) for $G_1$, since no multi-in-degree vertex except $D$ can be found above stage $l$ in both $G$ and $G_1$.☐

***Claim 4 (Picking out σ-paths).*** *If $\langle y, D, L\rangle \in (R(\langle a, b, h\rangle) \text{ of } G)$ ($\langle a, b, h\rangle \in E$, $\langle y, D, L\rangle \in E$, $1 < h \leq l$, $\langle u_i, v, l\rangle \in E$), there exists a σ-path that traverses both $\langle a, b, h\rangle$ and $\langle y, D, L\rangle$ in $G$.*

Claim 4 serves as a key tool to help us “split” the system invariant $ES1_{sub}$ for our constructed “smaller” graph. Claim 4 can be broken down to the following sub-claims (Claim 4a,4b,4c), depending on the varied locations of the edge $\langle a, b, h\rangle$. The key to its proof is that, the computed ($\boldsymbol{A}$ of $G$) by Operator 4 for $\langle w, D, L\rangle \in (R(e) \text{ of } G)$ fixes both $e$ and $\langle y, D, L\rangle$, and hence can be utilized for building ($ES1_{sub}$ of $G_1$) and infer the desired $\sigma$-path by our mathematical induction on $f(G)$.

***Claim 4a.*** *Given $\langle a, b, h\rangle \in E$ ($1 < h < l - 1$), $\langle w, D, L\rangle \in v - \cdots - D \subseteq E$. If $\langle w, D, L\rangle \in (R(\langle a, b, h\rangle) \text{ of } G)$, there exists a σ-path that traverses both $\langle a, b, h\rangle$ and $\langle w, D, L\rangle$ in $G$.*

*Proof.* Let $e = \langle a, b, h\rangle$.

The path from $v$ to $w$ in $G$ is uniquely determined, by the structure of $G$. Let it be denoted by $v - t - \cdots - r - w$. (The definition of paths $u_i - v_i - t_i - \cdots - r_i - w_i - D \subseteq E_1$ $(i = 1,2)$ might introduce a slight notational overlap about the “$t_i$,...,$r_i$,$w_i$” on the paths and those “$t$,...,$r$,$w$” on an arbitrarily designated $v - t - \cdots - r - w \subseteq E$. There is actually no direct correspondence between them, although the same letters “$t$,...,$r$,$w$” are shared.)

First note the following facts:

- By the sets ($\boldsymbol{A}$ of $G$) $\neq \emptyset$, ($\boldsymbol{B}$ of $G$) $\neq \emptyset$ computed for deciding “$\langle w, D, L\rangle \in (R(e) \text{ of } G)$” by Operator 4 in step 2 of $ZH\backslash step4$, it can be inferred that $v - t - \cdots - r - w - D \subseteq \boldsymbol{A}$ is a $\omega$-path.
- It also can be observed that, there exists a non-empty set $J \subseteq \{1,2\}$, such that for each $j \in J$: (i) $\langle u_j, v, l\rangle \in \big((\boldsymbol{A} \cap R(e)) \text{ of } G\big)$, because $(R(e) \text{ of } G)$ must contain a reachable path which traverses $\langle w, D, L\rangle$ via the vertex $v$ when applying Operator 3 for computing ($\boldsymbol{A}$ of $G$); (ii) $\langle w, D, L\rangle \in$ $\big(R(\langle u_j, v, l\rangle)$ of $G\big)$, because $\langle u_j, v, l\rangle \in (\boldsymbol{A}$ of $G)$ and $\langle w, D, L\rangle$ is the unique edge of stage $L$ in ($\boldsymbol{A}$ of $G$).

- The radical expansion forces each label on $v_j - t_j - \cdots - r_j - w_j - D$ $(j \in J)$ in $G_1$ to contain $(\lambda(v_j)[1{:}\,l-1]$ of $G_1) = \Big(\big(\lambda(u_j) \cap \lambda(v)\big)[1{:}\,l-1]$ of $G\Big) = \Big(\big(\lambda(u_j) \cap \lambda(v_j)\big)[1{:}\,l-1]$ of $G_1\Big)$ as a subset. Subsequently, "$\langle u_j, v, l\rangle \in (R(e)$ of $G)$" implies "$\langle u_j, v_j, l\rangle \in (R(e)$ of $G_1)$" (by the radical expansion of $\big(\lambda(v_j)$ of $G_1\big)$) and hence "$v_j - t_j - \cdots - r_j - w_j - D \subseteq (R(e)$ of $G_1)$" (by the radical expansion of $\big(\lambda(t_j)$ of $G_1\big)$,...,$\big(\lambda(w_j)$ of $G_1\big)$); "$\langle w, D, L\rangle \in (R(\langle u_j, v, l\rangle)$ of $G)$" implies "$\langle w_j, D, L\rangle \in$ $\big(R(\langle u_j, v_j, l\rangle)$ of $G_1\big)$" and hence "$v_j - t_j - \cdots - r_j - w_j - D \subseteq (R(\langle u_j, v_j, l\rangle)$ of $G_1)$". For detailed argument, if needed, see the renaming rules and the "transit" technique discussed in Appendix A.8.1 for ($R(E)$ of $G_1$).

Returning to the computation performed on $G$ which decides "$\langle w, D, L\rangle \in (R(e)$ of $G)$". Guided by the corresponding edge sets ($\boldsymbol{A}$ of $G$) and ($\boldsymbol{B}$ of $G$) that is involved in Operator 4, an appropriate $ES1_{sub}$ for the "smaller" graph $G_1$ can be constructed as follows.

If we choose in a "single-plank bridge" way that

$$(ES_temp \text{ of } G_1) =_{def} \left(\begin{array}{c}\left((\boldsymbol{A}[1{:}\,l-1] \text{ of } G) \cup \left\{e' \,\middle|\, \begin{array}{c} e' \in u_i - v_i - \cdots - w_i - D \subseteq E_1, \\ \langle u_i, v, l\rangle \in (\boldsymbol{A} \text{ of } G), i \in \{1,2\}\end{array}\right\}\right) \\ -\{e' \in E \mid e' \neq e \text{ is an edge of stage } h\}\end{array}\right), \tag{12}$$

then we still have $\big(\chi^D_{R(E)}(ES_temp)$ of $G_1\big) \neq \emptyset$, despite the removal of edges at stage $h$. That is because the computation of $\big(\chi^D_{R(E)}(ES_temp)$ of $G_1\big)$ is essentially the same as that of $\big(\psi_{R(E)-\{R(e)\}}\big(R(e)\big)$ of $G\big)$ when deciding "$\langle w, D, L\rangle \in (R(e)$ of $G)$". This is straightforward by the radical expansion—according to the above discussion, for each $\varepsilon \in \left(\left(\boldsymbol{A} - \left\{e' \in E \,\middle|\, \begin{array}{c} e' \neq e \text{ is an} \\ \text{edge of stage } h\end{array}\right\}\right) \text{ of } G\right)$ ( $\varepsilon \neq \langle w, D, L\rangle$ ), there exists $\hat{\varepsilon} \in E_1$ such that $\langle w_i, D, L\rangle \in$ $(R(\hat{\varepsilon})$ of $G_1)$ (where $i \in \{1,2\}$ such that $\langle u_i, v, l\rangle \in (\boldsymbol{A}$ of $G)$). As illustrated in Figure 8. Refer to Appendix A.8.2 for more detailed discussion, if needed.

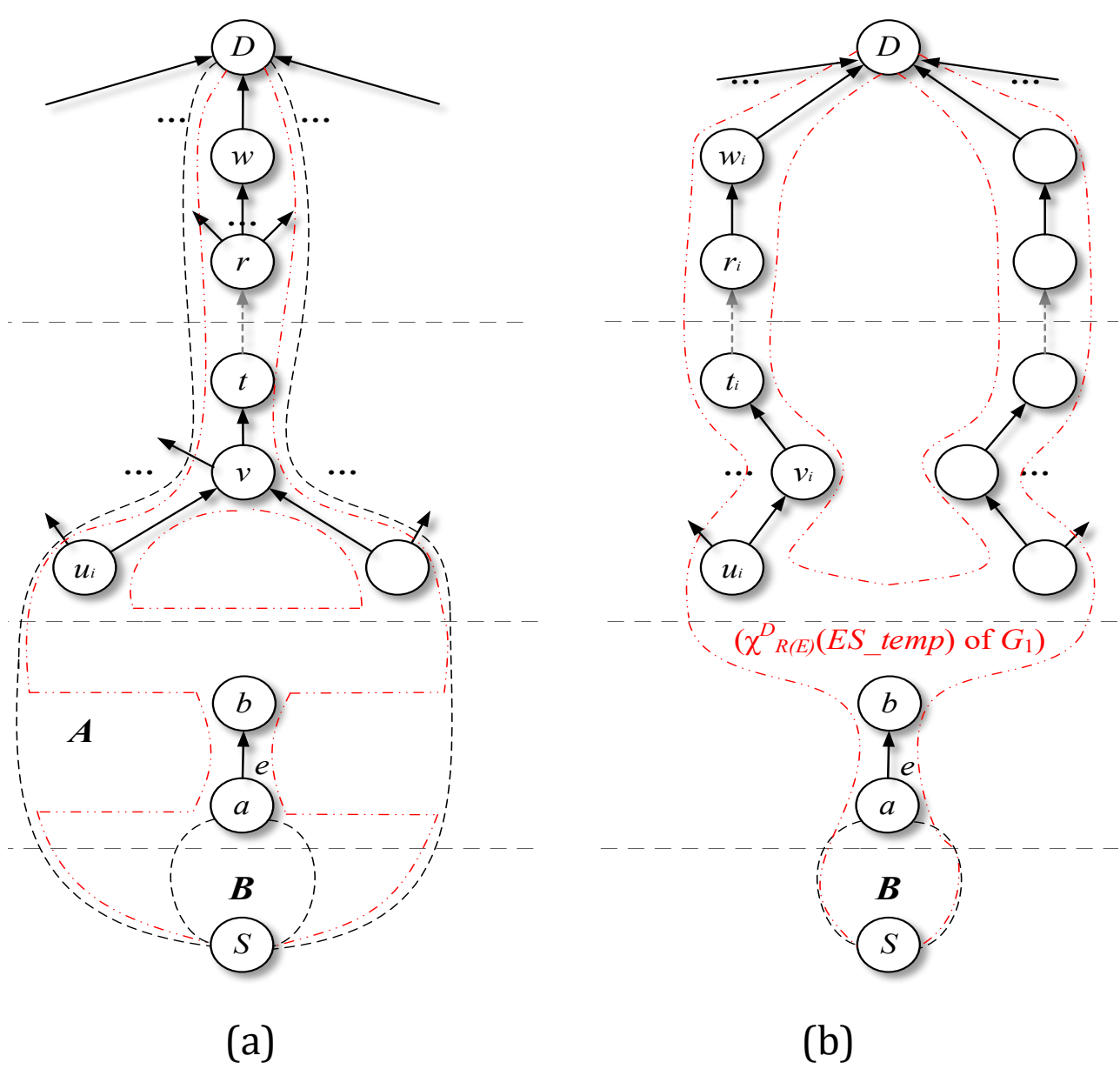

Figure 8: Illustration of Claim 4a

Since $\left(\chi^D_{R(E)}(ES_temp)\text{ of }G_1\right) \neq \emptyset$, by the definition of Operator 3, there must exist some $\langle \alpha, \beta, 2\rangle \in \left(\chi^D_{R(E)}(ES_temp)\text{ of }G_1\right)$ such that $H = \beta - \cdots - D \subseteq \left(\left[R(\langle \alpha, \beta, 2\rangle) \cap \chi^D_{R(E)}(ES_temp)\right]^D_\beta \text{ of } G_1\right) \neq \emptyset$. Assume $\langle u_1, v_1, l\rangle \in H$. Additionally, if $\left|\left(\chi^D_{R(E)}(ES_temp)[L{:}L]\text{ of }G_1\right)\right| = 2$, there must exist a path $H' = x - \cdots - D \subseteq \left(\chi^D_{R(E)}(ES_temp)\text{ of }G_1\right)$ such that $x$ is the unique non-sink vertex on both $H$ and $H'$; otherwise, assume $H' = \emptyset$.

Then, we can choose

$$(ES2\text{ of }G_1) =_{def} \left(\chi^D_{R(E)}(ES_temp)[L{:}L]\text{ of }G_1\right), \tag{13}$$

$$(ES1_{sub}\text{ of }G_1) =_{def} (S - \alpha - \beta) \cup H \cup H'. \tag{14}$$

If we define $(\lambda_{sub}(w_i)\text{ of }G_1) =_{def} (ES1_{sub}\text{ of }G_1) \cap (\lambda(w_i)[1{:}1] \cup \{e \in \lambda(w_i)[2{:}L] | \langle w_i, D, L\rangle \in R(e)\})$ $(i = 1,2)$, then $(ES1_{sub}\text{ of }G_1)$ obeys criteria (i),(ii),(iv). Criterion (iii) is apparently obeyed, since we can assume that no $\sigma$-path contained in $(ES1_{sub}\text{ of }G_1)$ exists, otherwise Claim 4a is proved. Criterion (v) is also obeyed, since $\langle w_1, D, L\rangle \in (R(e)\text{ of }G_1)$ for some $e \in [(S - \alpha - \beta) \cup H]^x_S$ while $\langle w_2, D, L\rangle \in (R(e')\text{ of }G_1)$ for some $e' \in [(S - \alpha - \beta) \cup H]^x_S - \{e\}$. Otherwise, we simply define $(ES2\text{ of }G_1) =_{def} \{\langle w_1, D, L\rangle\}$ and $(ES1_{sub}\text{ of }G_1) =_{def} (S - \alpha - \beta) \cup H$. Meanwhile, we can assume that no $\sigma$-path contained in $(ES1_{sub}\text{ of }G_1)$ exists, and hence $(\lambda_{sub}(w_i)\text{ of }G_1)$ $(i = 1,2)$ has no obligation to be set to include the whole $[(S - \alpha - \beta) \cup H]^x_S$.

It can be further straightforwardly observed that $([ES1_{sub}]^D_S[L{:}L]\text{ of }G_1) = (ES2\text{ of }G_1) \neq \emptyset$, by the definition of $(ES_temp\text{ of }G_1)$ and by the fact that $\left[\left(\chi^D_{R(E)}(ES_temp)\text{ of }G_1\right)\right]^D_S[L{:}L] \subseteq \left\{\begin{matrix}\langle w_1, D, L\rangle, \\ \langle w_2, D, L\rangle\end{matrix}\right\}$.

Then, by our mathematical induction hypothesis (H1), via the PA algorithm, we can infer that there exists some $\sigma$-path $SP = S - \cdots - a - b - \cdots - u_i - v_i - \cdots - w_i - D \subseteq (ES1_{sub}\text{ of }G_1)$ $(i \in \{1,2\})$ in $G_1$ by the PA algorithm.

Note that $SP[1: l-1] \cup (u_i - v - \cdots - w - D) \subseteq (\boldsymbol{A} \text{ of } G)$, and hence straightforwardly $SP[1: l-1] \cup (u_i - v - \cdots - w - D)$ is a $\sigma$-path in $G$ by Operator 3.☐

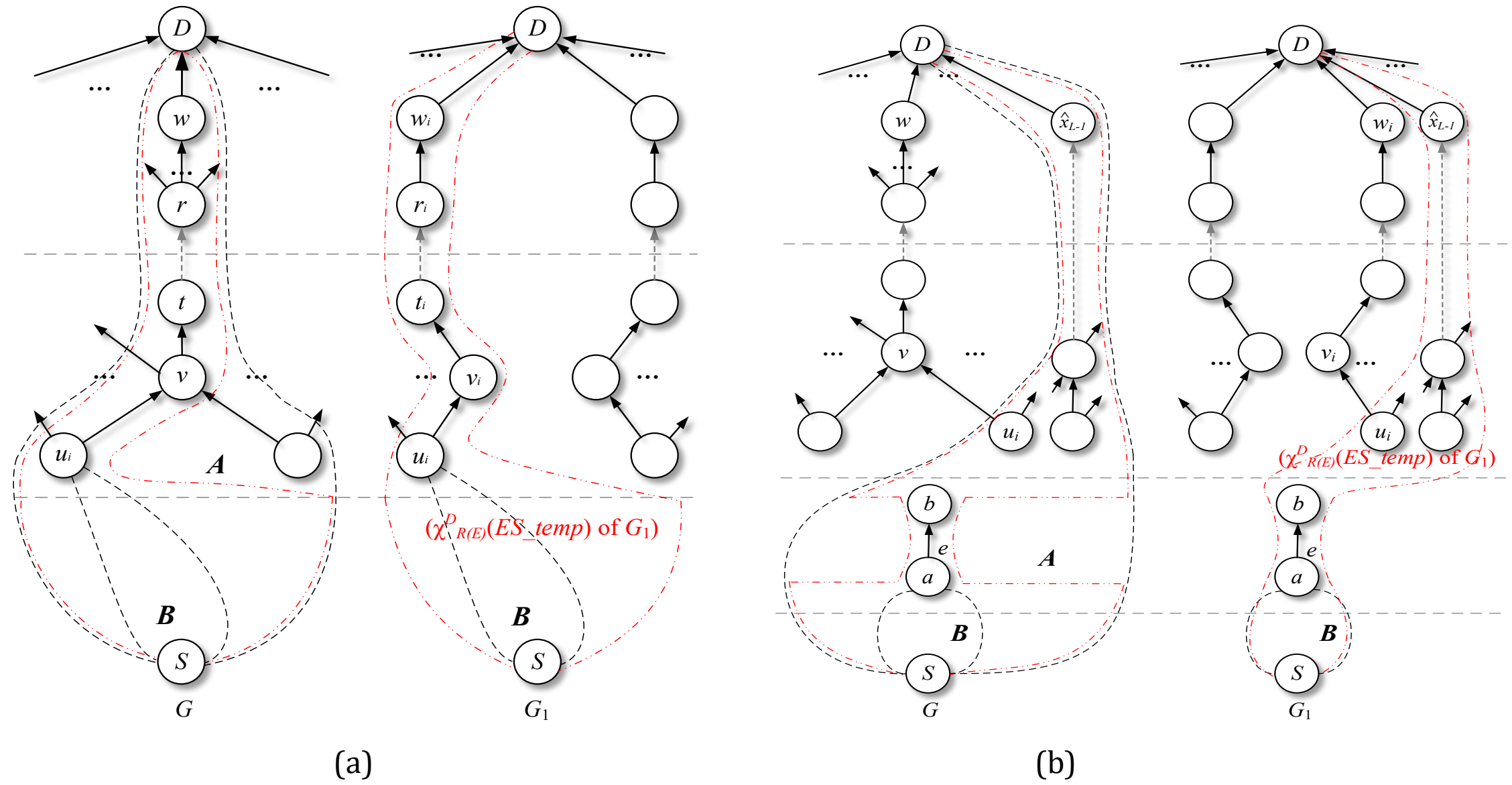

Figure 9: Illustration of Claim 4b, Claim 4c

***Claim 4b.*** *Given $\langle a, b, h\rangle \in E$ ($h > 1$, $l - 1 \leq h \leq l$), $\langle w, D, L\rangle \in v - \cdots - D \subseteq E$. If $\langle w, D, L\rangle \in (R(\langle a, b, h\rangle) \text{ of } G)$, there exists a $\sigma$-path that traverses both $\langle a, b, h\rangle$ and $\langle w, D, L\rangle$ in $G$.*

***Claim 4c.*** *Given $\langle a, b, h\rangle \in E$ ($1 < h \leq l$), $\langle \hat{x}_{L-1}, D, L\rangle \notin v - \cdots - D \subseteq E$. If $\langle \hat{x}_{L-1}, D, L\rangle \in (R(\langle a, b, h\rangle) \text{ of } G)$, there exists a $\sigma$-path that traverses both $\langle a, b, h\rangle$ and $\langle \hat{x}_{L-1}, D, L\rangle$ in $G$.*

Claim 4b,4c are similar to Claim 4a (each illustrated in Figure 9(a) and Figure 9(b)), despite that we can simply choose $H' = \emptyset$, because $|(\boldsymbol{A}[l: l] \text{ of } G)| = 1$ for Claim 4b and $\langle a, b, h\rangle \in (\boldsymbol{A} \text{ of } G) \subseteq (\boldsymbol{A} \text{ of } G_1)$ for Claim 4c.

*A.6.2 The construction of a mathematical equivalent $G_2$ based on $G_1$, the definition of $(ES1_{sub} \text{ of } G_2)$, and the proof of Claim 5*

***Claim 5.*** *If $([ES1_{sub}]^D_S[L: L] \text{ of } G) = (ES2 \text{ of } G)$, then $\left(\forall \langle w, D, L\rangle \in ([ES1_{sub}]^D_S \text{ of } G)\right)$ $(\exists\, \sigma - path\ S - \cdots - w - D \subseteq (ES1_{sub} \text{ of } G))$.*

Based on the following Step I, II and III, Claim 5 can get proved.

*A.6.2.1 Step I: The construction of mathematical equivalent $G_2$ based on $G_1$*

Our inductive proving framework requires finding proper $(ES1_{sub} \text{ of } G_1)$ which is mathematically equivalent to the provided $(ES1_{sub} \text{ of } G)$. To do that, a graph $G_2$ is further constructed based on $G_1$. In $G_2$, with the help of Claim 4, we precisely list out each $\sigma$-path in $G$ in a flavor of "mathematical analysis", as follows.

By Claim 4 and criterion (ii) (that is $(\lambda_{sub}(y) \text{ of } G) \subseteq ((\lambda(y)[1: 1] \cup \{e \in \lambda(y)[2: L] | \langle y, D, L\rangle \in R(e)\}) \text{ of } G)$), for each $\langle w, D, L\rangle \in E$, if $e \in ((\lambda(w)[1: 1] \cup \{e \in \lambda(w)[2: L] | \langle y, D, L\rangle \in R(e)\}) \text{ of } G)$ ($e = \langle a, b, h\rangle \in E$, $1 < h \leq l$), there should exist at least one $\sigma$-path that traverses both $e$ and $\langle w, D, L\rangle$ in $G$. Just pick one such $\sigma$-path for each pair of $e$ and $\langle w, D, L\rangle$. If the $\sigma$-path traverses $\langle u_i, v, l\rangle$ for some $i \in \{1,2\}$, then we can denote it by $P^{\langle e|i|w\rangle}$; otherwise, just denote it by $P^{\langle e|0|w\rangle}$.

Then, for each such $P^{\langle e|i|w\rangle}$ (where $e = \langle a, b, h\rangle, 1 < h \leq l, i \in \{0,1,2\}$) in $G$, we introduce a new path $x_{L-2}^{(e|i|w)} - x_{L-1}^{(e|i|w)} - D$ to $G_1$, such that there will exist a path $X^{\langle e|i|w\rangle}$ as follows in the resulted graph $G_2$:

(i) For $i \in \{1,2\}$: let $X^{\langle e|i|w\rangle} =_{def} \left(u_i - v_i - x_{l+1}^{(e|i|w)} - \cdots - x_{L-2}^{(e|i|w)}\right) \cup \left(x_{L-2}^{(e|i|w)} - x_{L-1}^{(e|i|w)} - D\right)$, where $u_i - v - \cdots - w - D \subseteq P^{\langle e|i|w\rangle} \subseteq E$ and $u_i - v_i - x_{l+1}^{(e|i|w)} - \cdots - x_{L-2}^{(e|i|w)} \subseteq u_i - v_i - \cdots - w_i - D \subseteq E_1$.

(ii) For $i = 0$: let $X^{\langle e|i|w\rangle} =_{def} \left(\hat{x}_{l-1} - \hat{x}_l - x_{l+1}^{(e|i|w)} - \cdots - x_{L-2}^{(e|i|w)}\right) \cup \left(x_{L-2}^{(e|i|w)} - x_{L-1}^{(e|i|w)} - D\right)$, where $\hat{x}_{l-1} - \hat{x}_l - x_{l+1}^{(e|i|w)} - \cdots - x_{L-2}^{(e|i|w)} \subseteq P^{\langle e|i|w\rangle} \subseteq E \cap E_1$.

Set $\lambda_2$ for $G_2$:

$$\left(\lambda\left(x_{L-1}^{(e|i|w)}\right) \text{ of } G_2\right) = P^{\langle e|i|w\rangle}[1: l-1] \cup X^{\langle e|i|w\rangle}[l: L-1] \text{ (where } X^{\langle e|i|w\rangle} \subseteq E_2\text{)},$$
$$(\lambda(D) \text{ of } G_2) = (\lambda(D) \text{ of } G_1) \cup \bigcup_{X^{\langle e|i|w\rangle} \subseteq E_2} \left(x_{L-2}^{(e|i|w)} - x_{L-1}^{(e|i|w)} - D\right). \tag{15}$$

All the other labels remain the same as they were defined in $G_1$. We thus get a new graph $G_2 =< \widehat{V_2}, E_2, S, D, L, \lambda_2 >$ (see Figure 6(c)). Note again that $\widehat{V_n}$ is used to distinguish from the notation "$V_n$" in Definition 1.

The above "singleton" definition of $\left(\lambda\left(x_{L-1}^{(e|i|w)}\right) \text{ of } G_2\right)$ ($X^{\langle e|i|w\rangle} \subseteq E_2$) makes the label be controlled to correspond to exactly one $\sigma$-path $P^{\langle e|i|w\rangle}[1: l-1] \cup X^{\langle e|i|w\rangle}$ in $G_2$.

For latter usage, as with each $\hat{x}_l$ at stage $l$ of $G_2$, we arbitrarily pick one path $X^{\langle e|i|w\rangle} \subseteq E_2$ (if exists, where $\hat{x}_l$ appears on $X^{\langle e|i|w\rangle}$, $i \in \{0,1,2\}$), and add one path "$x_{L-2}^{(e|i|w)} - \omega_{L-1}^{\hat{x}_l} - D$" to $G_2$, such that there will exist one path $Pspare_{\hat{x}_l} =_{def} \left(\hat{x}_l - \omega_{l+1}^{\hat{x}_l} - \cdots - \omega_{L-1}^{\hat{x}_l} - D\right) = X^{\langle e|i|w\rangle}[l+1: L-2] \cup \left(x_{L-2}^{(e|i|w)} - \omega_{L-1}^{\hat{x}_l} - D\right) = \left(\hat{x}_l - x_{l+1}^{(e|i|w)} - \cdots - x_{L-2}^{(e|i|w)}\right) \cup \left(x_{L-2}^{(e|i|w)} - \omega_{L-1}^{\hat{x}_l} - D\right)$. Define the labels as:

$$\left(\lambda\left(\omega_{L-1}^{\hat{x}_l}\right) \text{ of } G_2\right) = \left(P^{\langle e|i|w\rangle}[1: l-1] \cup X^{\langle e|i|w\rangle}[l: l] \cup Pspare_{\hat{x}_l}[l+1: L-1]\right),$$
$$(\lambda(D) \text{ of } G_2) = (\lambda(D) \text{ of } G_2) \cup Pspare_{\hat{x}_l}. \tag{16}$$

Note that $P^{\langle e|i|w\rangle}[1: l-1] \cup X^{\langle e|i|w\rangle}[l: l] \cup Pspare_{\hat{x}_l}$ is a $\sigma$-path, which is also exactly contained by $\left(\lambda\left(\omega_{L-1}^{\hat{x}_l}\right) \text{ of } G_2\right)$.

The number of the above newly introduced paths for all possible combinations of $e$,$w$ and $i$ is a polynomial in $|E|$.

We thus complete the construction of $G_2$.

It can be easily obtained again that:

***Remark 1 (Claim 1 for $\mathbf{G_2}$).*** *$f(G_2) < f(G)$.*

***Remark 2 (Claim 2 for $\mathbf{G_2}$).*** *If $\left(\chi_{R(E)}^{D}(\lambda(D)) \text{ of } G\right) \neq \emptyset$, we have:*

$$(ES1\ of\ G_2) \supseteq \begin{pmatrix} \left((ES1\ of\ G) - \left\{e \middle| e \in u_i - v - \cdots - D \subseteq E, i \in \{1,2\}\right\}\right) \\ \cup \left\{e \middle| \begin{matrix} e \in u_i - v_i - \cdots - w_i - D \subseteq E_1, \\ (R(\langle u_i, v, l\rangle)\ of\ G) \neq \emptyset, i \in \{1,2\} \end{matrix}\right\} \\ \cup \left\{\hat{e} \middle| \begin{matrix} \hat{e} \in X^{\langle e|i|w\rangle} \subseteq E_2, \\ i \in \{0,1,2\} \end{matrix}\right\} \cup \left\{\hat{e} \middle| \hat{e} \in Pspare_{\hat{x}_l} \subseteq E_2\right\} \end{pmatrix} \neq \emptyset. \tag{17}$$

*And hence,* $\left(\chi^D_{R(E)}\left(\lambda(D)\right)\ of\ G\right) \neq \emptyset$ *implies* $(ES1\ of\ G_2) = \left(\chi^D_{R(E)}\left(\lambda(D)\right)\ of\ G_2\right) \neq \emptyset$*.*

*Proof.* This is clear. The renaming rules for $(R(E)$ of $G_2)$ are analogous to those for $(R(E)$ of $G_1)$, since $G_2$ only adds some $\omega$-paths to $G_1$. Moreover, each of those "$X^{\langle e|i|w\rangle}$" is on some $\sigma$-path $P$ in $G_2$. For any $e = \langle a, b, k\rangle \in P$ $(0 < k < L)$, $(R(e)$ of $G_2)$ contains $P[k+1 : L]$ and further contains $\langle w, D, L\rangle$. Hence they can all be kept in $\left(\chi^D_{R(E)}\left(\lambda(D)\right)\right.$ of $\left.G_2\right)$.☐

***Remark 3 (Claim 3 for*** $G_2$***).*** *$G_2$ satisfies Definition 3(b).*

*Proof.* Definition 3(b) is clearly fulfilled, because: (i) we didn't change the in-degrees of other vertices except $D$ in $G_1$; (ii) the vertices on those newly introduced "$X^{\langle e|i|w\rangle}[l+1: L-1]$" are single-in-degree vertices.☐

*A.6.2.2 Step III: The definition of* $(ES1_{sub}\ of\ G_2)$

*A.6.2.2.1 Step III(a): The initial definition of* $(ES1_{sub}\ of\ G_2)$

Guided by the given $(ES2$ of $G)$, $(\lambda_{sub}(w)$ of $G)$ and $(ES1_{sub}$ of $G)$, we can define an initial $(ES1_{sub}$ of $G_2)$ as follows:

- $(ES2$ of $G_2) =_{\text{def}} \left\{\hat{e} \middle| \hat{e} \in X^{\langle e|i|w\rangle}[L: L] \subseteq E_2, \langle w, D, L\rangle \in (ES2 \text{ of } G)\right\}$.
- $(ES1_{sub}[l+1: L]$ of $G_2) =_{\text{def}} \left\{\hat{e} \middle| \hat{e} \in X^{\langle e|i|w\rangle}[l+1: L], \langle x^{(e|i|w)}_{L-1}, D, L\rangle \in (ES2 \text{ of } G_2)\right\}$.

  *(Substitute "$P^{\langle e|i|w\rangle}[l+1: L]$" by "$X^{\langle e|i|w\rangle}[l+1: L]$" in $G_2$, if $\langle w, D, L\rangle \in (ES1_{sub}\ of\ G)$.)*
- $(ES1_{sub}[1: l]$ of $G_2) =_{\text{def}} ([ES1_{sub}]^D_S[1: l]$ of $G)$.
- $\left(\lambda_{sub}\left(x^{(e|i|w)}_{L-1}\right) \text{ of } G_2\right) =_{\text{def}}$

$$\begin{cases} \begin{pmatrix} ([ES1_{sub}]^D_S \text{ of } G) \cap \\ (\lambda_{sub}(w) \text{ of } G) \cap \\ \left(\lambda\left(x^{(e|i|w)}_{L-1}\right)[1: L-2] \text{ of } G_2\right) \end{pmatrix} \cup \left\{\langle x^{(e|i|w)}_{L-2}, x^{(e|i|w)}_{L-1}, L-1\rangle\right\}, & i = 0 \\ \begin{pmatrix} ([ES1_{sub}]^D_S \text{ of } G) \cap \\ (\lambda_{sub}(w) \text{ of } G) \cap \\ \left(\lambda\left(x^{(e|i|w)}_{L-1}\right)[1: l] \text{ of } G_2\right) \end{pmatrix} \cup X^{\langle e|i|w\rangle}[l+1: L-1], & i \in \{1,2\} \end{cases}$$

  where $\langle x^{(e|i|w)}_{L-1}, D, L\rangle \in (ES2$ of $G_2)$.

  *(Partition the set "$([ES1_{sub}]^D_S\ of\ G)$" by "$\left(\lambda\left(x^{(e|i|w)}_{L-1}\right)\ of\ G_2\right)$".)*

A mild abuse of notation is introduced here for readability. For instance, when defining $(ES1_{sub}[1: l]$ of $G_2)$, we simply write " $([ES1_{sub}]^D_S[1: l]$ of $G)$ " instead of precisely writing "$\bigcup_{e \in ([ES1_{sub}]^D_S[1:l] \text{ of } G)} \left\{\hat{e} \middle| \begin{matrix} \hat{e} = \langle u_i, v_i, l\rangle \text{ if } e = \langle u_i, v, l\rangle; \\ \hat{e} = e \text{ otherwise. } (i \in \{1,2\}) \end{matrix}\right\}$".

Explain the motivation of the construction as follows.

If $(ES1_{sub}$ of $G)$ contains $\langle w, D, L\rangle \in E$, we can use all related "$X^{\langle e|i|w\rangle}[l+1:L]$" in $G_2$ to substitute "$P^{\langle e|i|w\rangle}[l+1:L]$". Then, we can exactly choose edges for $\left(\bigcup_{\langle y,D,L\rangle \in ES2} \lambda_{sub}(y) \text{ of } G_2\right)[1:l]$, such that $\left(\bigcup_{\langle y,D,L\rangle \in ES2} \lambda_{sub}(y) \text{ of } G_2\right)[1:l]$ is essentially the same as $\left(\bigcup_{\langle w,D,L\rangle \in ES2} \lambda_{sub}(w) \text{ of } G\right)[1:l]$. Consequently, different from the computation of $([ES1_{sub}]_S^D \text{ of } G)$, the computation of $([ES1_{sub}]_S^D \text{ of } G_2)$ will use all those related "$X^{\langle e|i|w\rangle}[l+1:L]$" instead of "$P^{\langle e|i|w\rangle}[l+1:L]$".

To maintain consistency with the criteria for the constitution of $(ES1_{sub}$ of $G_2)$, the set $\left(\lambda_{sub}\left(x_{L-1}^{(e|i|w)}\right) \text{ of } G_2\right)$ $(\langle x_{L-1}^{(e|i|w)}, D, L\rangle \in (ES2 \text{ of } G_2))$ is accordingly defined, by partitioning the set $([ES1_{sub}]_S^D \text{ of } G)$ using each $\sigma$-path "$P^{\langle e|i|w\rangle}[1:l-1] \cup X^{\langle e|i|w\rangle}[l:L]$" specified by "$\left(\lambda\left(x_{L-1}^{(e|i|w)}\right) \text{ of } G_2\right)$".

The benefit of this construction lies in that, as will be certificated later, the usage of "$X^{\langle e|i|w\rangle}[l+1:L]$" and the "singleton" definition of "$\left(\lambda\left(x_{L-1}^{(e|i|w)}\right) \text{ of } G_2\right)$" will make it easier and clearer to "recover" those $\sigma$-paths "$P^{\langle e|i|w\rangle}$"$\subseteq$ $(ES1_{sub}$ of $G)$ from those $\sigma$-paths "$P^{\langle e|i|w\rangle}[1:l-1] \cup X^{\langle e|i|w\rangle}$"$\subseteq$ $(ES1_{sub}$ of $G_2)$.

Now we finish the explanation of motivation.

The key that enables the above partition lies in: (1) $\left(\lambda\left(x_{L-1}^{(e|i|w)}\right) \text{ of } G_2\right) = \left(\lambda\left(x_{L-1}^{(e|i|w)}\right)[1:1] \text{ of } G_2\right) \cup \left\{e \,\middle|\, \langle x_{L-1}^{(e|i|w)}, D, L\rangle \in (R(e) \text{ of } G_2)\right\}$; (2) $\left(\lambda\left(x_{L-1}^{(e|i|w)}\right) \text{ of } G_2\right)$ contains only one σ-path $P$. Hence for any $e = \langle a, b, k\rangle \in P$ $(0 < k < L)$, $(R(e)$ of $G_2)$ contains $P[k+1:L]$ and further contains $\langle w, D, L\rangle$.

Apparently, $(ES1_{sub} \text{ of } G_2) = \left(\left(ES2 \cup \left(\left(\bigcup_{\langle y,D,L\rangle \in ES2} \lambda_{sub}(y)\right) \cap ES1\right)\right) \text{ of } G_2\right)$ and $(\lambda_{sub}(y) \text{ of } G_2) \subseteq \left((\lambda(y)[1:1] \cup \{e \in \lambda(y)[2:L] | \langle y, D, L\rangle \in R(e)\}) \text{ of } G_2\right)$ $(\langle y, D, L\rangle \in E_2)$, and hence criteria (i),(ii) are fulfilled. Criterion (iii) is not violated, because we did not add $\sigma$-paths which traverse edges in $(ES2$ of $G_2)$. Criterion (iv) is obeyed, because: $(ES1_{sub}$ of $G_2)$ and $(\lambda_{sub}(y)$ of $G_2)$ both choose edges accurately, and thus $(ES1_{sub}$ of $G_2)$ never add edges which do not appear in $(ES1_{sub}$ of $G)$; besides, there exists no multi-in-degree vertex except $D$ in $G_2$ above stage $l$.

As for criterion (v): if $v$ of stage $l$ does not appear on the said $S - \cdots - a_i - \cdots - D \subseteq ([ES1_{sub}]_S^D \text{ of } G)$ by criterion (v) or it just appears with $i < l$, then there exists $S - a_1 - \cdots - a_i \subseteq ([ES1_{sub}]_S^D \text{ of } G_2)$ such that we still have that "$([ES1_{sub}]_S^D \text{ of } G_2)$ contains one $\langle a_j, *, j+1\rangle$ at most for each $a_j$ $(1 \leq j < i)$ while two $\langle a_i, *, i+1\rangle$ at least, and $S - a_1 - \cdots - a_i \nsubseteq (\lambda_{sub}(y) \text{ of } G_2)$ for $\langle y, D, L\rangle \in (ES2 \text{ of } G_2)$"; if $v = a_l$, $i \geq l$ and $a_l$ appears on the said $S - \cdots - a_i - \cdots - D \subseteq ([ES1_{sub}]_S^D \text{ of } G)$, then there exists some $P = S - \cdots - v_1 - \cdots - t_1 \subseteq E_2$ (where $t_1$ lies at stage $L-2$) and $P[1:l] \nsubseteq (\lambda_{sub}(y) \text{ of } G_2)$ for $\langle y, D, L\rangle \in (ES2 \text{ of } G_2)$. Thus, it's plain to obtain:

***Remark 4 (Initial $(\boldsymbol{ES1_{sub}}$ of $\boldsymbol{G_2})$, on the constitution).*** *The initially defined $(ES1_{sub}$ of $G_2)$ and $\left(\lambda_{sub}\left(x_{L-1}^{(e|i|w)}\right) \text{ of } G_2\right)$ $(\langle x_{L-1}^{(e|i|w)}, D, L\rangle \in (ES2 \text{ of } G_2))$ satisfy the criteria (i),(ii),(iii),(iv),(v) defined in PA algorithm.*

*A.6.2.2.2 Step III(b): The compensation to $(ES1_{sub}\ of\ G_2)$ to prove Claim 5*

For the initially defined $(ES1_{sub}\text{ of }G_2)$, an issue might arise regarding the connectivity prerequisite $([ES1_{sub}]_S^D[L{:}L]\text{ of }G_2) = (ES2\text{ of }G_2)$, due to the "split" of the multi-in-degree vertex $v$. Further "**compensation**" to $(ES1_{sub}\text{ of }G_2)$ should be done conditionally, to keep $(ES1_{sub}\text{ of }G_2)$ mathematically equivalent to the provided $(ES1_{sub}\text{ of }G)$ and to maintain the connectivity prerequisite. The investigation of such $(ES1_{sub}\text{ of }G_2)$ is the deepest and core discussion of the entire paper. This method (i.e., the existence of this compensation, which might have not ever been established by all previous studies), we think, just has the same logical power with the seek and construction of uncomputable functions (Church, 1936; Turing, 1936) by the method of diagonalization.

**Case 1.** $\left\{\begin{matrix}\langle u_1, v, l\rangle,\\ \langle u_2, v, l\rangle\end{matrix}\right\} \cap ([ES1_{sub}]_S^D\text{ of }G) = \{\langle u_1, v, l\rangle | (R(\langle u_2, v, l\rangle)\text{ of }G) \neq \emptyset\}$. (The discussion is symmetrical if choosing $\langle u_2, v, l\rangle$.)

Note that $(R(\langle u_2, v, l\rangle)\text{ of }G) \neq \emptyset$ implies that there will exist some $\langle x_{L-1}^{(e|2|w)}, D, L\rangle \in (ES1\text{ of }G_2)$.

The homomorphic compensation is done constructively, as follows. The idea is to find a proper path to repair the connectivity prerequisite if $\langle u_2, v, l\rangle$ does not appear in $(ES1_{sub}\text{ of }G_2)$.

**Case 1a.** $\left|([ES1_{sub}]_S^D[L{:}L]\text{ of }G)\right| > 1$.

In this case, there exists $S - a_1 - \cdots - a_i \subseteq ([ES1_{sub}]_S^D\text{ of }G)$ $(i > 1)$ such that $([ES1_{sub}]_S^D\text{ of }G)$ contains one $\langle a_j, *, j+1\rangle$ at most for each $a_j$ $(1 \leq j < i)$ while two $\langle a_i, *, i+1\rangle$ at least, and $S - a_1 - \cdots - a_i \nsubseteq (\lambda_{sub}(y)\text{ of }G)$ for $\langle y, D, L\rangle \in (ES2\text{ of }G)$.

Pick a path $P \subseteq (ES1_{sub}\text{ of }G)$ that contains $S - a_1 - \cdots - a_i$; Since $\langle a_{i-1}, a_i, i\rangle \in (ES1_{sub}\text{ of }G)$, a $\sigma$-path that contains $\langle a_{i-1}, a_i, i\rangle$ must exist in $G$ according to Claim 4. This $\sigma$-path must be listed in $G_2$. Hence correspondingly, there exists $P' \subseteq (ES1_{sub}\text{ of }G_2)$ that contains $S - a_1 - \cdots - a_i$ too (if $a_j = v$ in $G$, then denote $a_j = v_1$ and let $a_{j+1}, \ldots, a_i$ be all vertices on $v_1 - \cdots - t_1$ in $G_2$, where $1 \leq j \leq i$ and $t_1$ is a vertex at stage $L-2$). In $G$, from $\langle u_2, v, l\rangle$ down to $S$, we seek for such a path: every time the path will reach to meet $P$, we choose another edge to continue our "downward searching" (Since $d^-(v) > 1$, we have $d^-(x) > 1$ for each path $x - \cdots - v \subseteq E$ that starts above stage 1 by item 3 of Definition 3(b)). Hence, search down until we meet any path $Q \subseteq ([ES1_{sub}]_S^D\text{ of }G)$ of a "branch" other than the one containing $P$ (hence $Q \cap \{\langle u_1, v, l\rangle, \langle u_2, v, l\rangle\} = \emptyset$), before reaching $S$; otherwise, we directly reach $S$ and just regard $Q = \emptyset$ might as well.

Let the vertex of intersection be $\xi$ and the traversed path be $H = \xi - \cdots - v$, as illustrated in Figure 10. Turn $T = [Q]_S^\xi \cup H[1{:}l-1] \cup \{\langle u_2, v_2, l\rangle\} \cup Pspare_{v_2}$ into a $\sigma$-path of $G_2$, by properly **expanding** the labels on $T$. **Add** $[T]_\xi^{v_2}$ to $(ES1_{sub}\text{ of }G_2)$, by setting $\left(\lambda\left(\omega_{L-1}^{v_2}\right)\text{ of }G_2\right) =_{\text{def}} T[1{:}L-1]$, $\left(\lambda_{sub}\left(\omega_{L-1}^{v_2}\right)\text{ of }G_2\right) =_{\text{def}} T[1{:}L-1]$ and $(ES2\text{ of }G_2) =_{\text{def}} (ES2\text{ of }G_2) \cup \left\{\langle \omega_{L-1}^{v_2}, D, L\rangle\right\}$. If $H$ intersects with other "subbranches" of the branch that contains $P$ —i.e., there exists some "bridge" $\beta - \cdots - \gamma \subseteq ([ES1_{sub}]_S^D\text{ of }G)$ such that (i) $\beta \in \cup_{0<i<l-1} V_i$ is on some "subbranch" of the "branch" that contains P and (ii) $\gamma \in \cup_{1<i<l} V_i$ is on $H$, **prune** the entire "bridge" from each $(\lambda_{sub}(y)\text{ of }G_2)$ $(\langle y, D, L\rangle \in (ES2\text{ of }G_2))$, so that $(ES1_{sub}\text{ of }G_2)$ obeys criteria (iv).

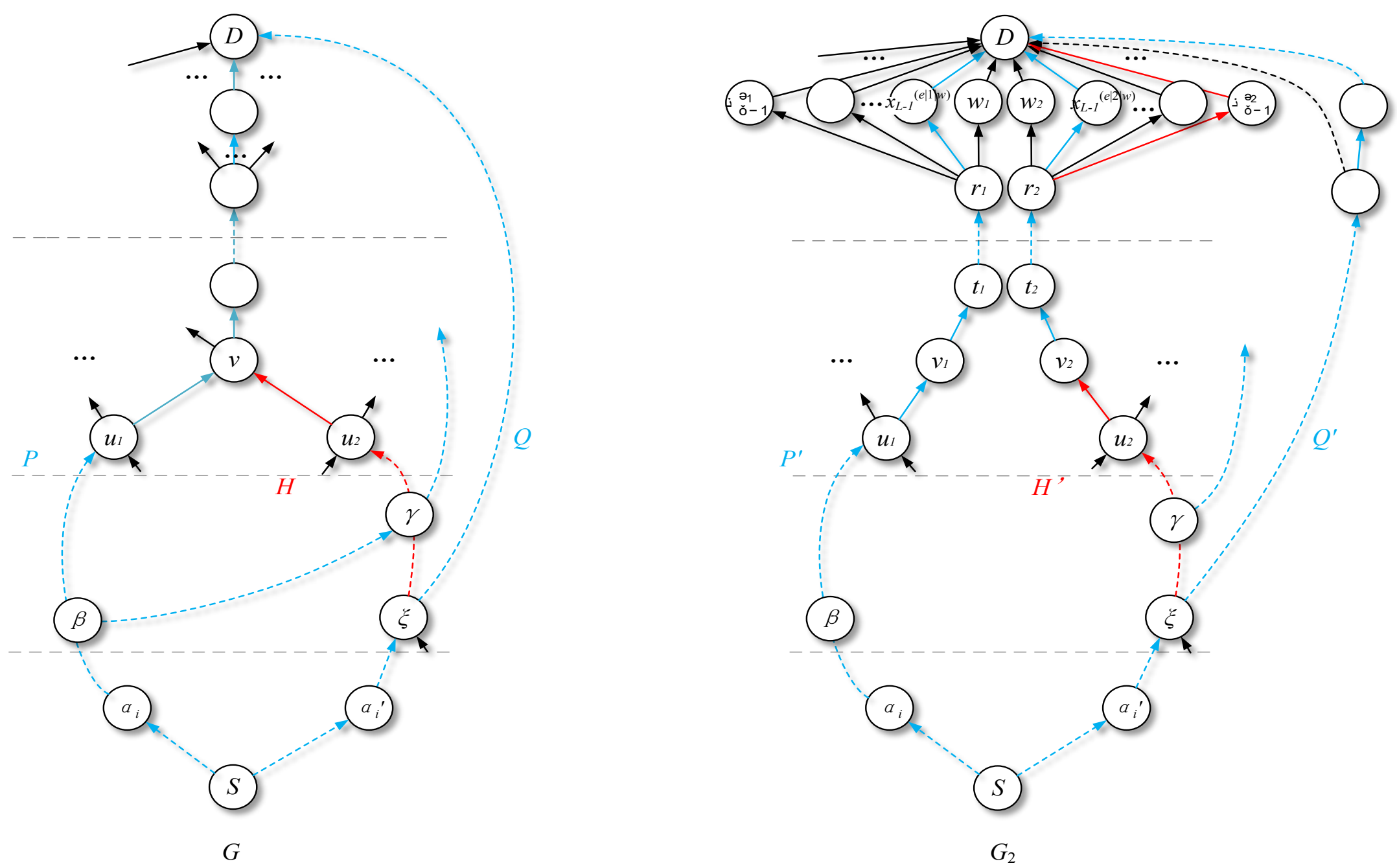

Figure 10: A typical case of the homomorphic compensation

Now, with the help of $[T]_{\xi}^{v_2}$, we have $([ES1_{sub}]_S^D[L{:}\,L]$ of $G_2) = (ES2$ of $G_2)$.— Each $X^{\langle e|2|w\rangle}[l+1{:}\,L]$ becomes a "subbranch" of $T$ and some "subbranches" of the branch that contains $P$ become "subbranches" of $T$ . Hence the pruning of "bridges" won't disturb the connectivity. We already expanded ($ES2$ of $G_2$) to include $\langle \omega_{L-1}^{v_2}, D, L\rangle$ and we also expanded ($ES1_{sub}$ of $G_2$) to include $T$ ($[T]_S^{\xi}$ is originally in ($ES1_{sub}$ of $G_2$)).

Note that we did not change the criteria (i),(ii) which ($[ES1_{sub}]_S^D$ of $G_2$) should obey. Criterion (iv) is obeyed since all possible "bridges" are pruned. Since each $\left(\lambda\left(x_{L-1}^{(e|i|w)}\right) \text{ of } G_2\right)$ is defined to exactly contain one $\sigma$-path, expanding labels on $T$ bring us no more $\sigma$-paths than $T$ itself. $T$ is the only newly introduced $\sigma$-path. We have set $\left(\lambda(\omega_{L-1}^{v_2}) \text{ of } G_2\right) = T[1{:}\,L-1]$ and $\left(\lambda_{sub}(\omega_{L-1}^{v_2}) \text{ of } G_2\right) = T[1{:}\,L-1]$. Thus, ($ES1_{sub}$ of $G_2$) obeys criterion (iii).

The discussion on the obedience of criterion (v) by the initial ($ES1_{sub}$ of $G_2$) still holds after compensation:

Noting the fact that $\xi$ is never on $S - a_1 - \cdots - a_i$ in $G$ , the fact that we did not change any "$\left(\lambda_{sub}\left(x_{L-1}^{(e|i|w)}\right) \text{ of } G_2\right)$" and the fact that $\left(\lambda_{sub}(\omega_{L-1}^{v_2}) \text{ of } G_2\right)$ only contains $T[1{:}\,L-1]$, if $v$ of stage $l$ does not appear on the said $S - \cdots - a_i - \cdots - D \subseteq ([ES1_{sub}]_S^D$ of $G)$ by criterion (v) or it just appears with $i < l$, then there exists $S - a_1 - \cdots - a_i \subseteq ([ES1_{sub}]_S^D$ of $G_2)$ such that we still have that "($[ES1_{sub}]_S^D$ of $G_2$) contains one $\langle a_j, *, j+1\rangle$ at most for each $a_j$ $(1 \leq j < i)$ while two $\langle a_i, *, i+1\rangle$ at least, and $S - a_1 - \cdots - a_i \nsubseteq (\lambda_{sub}(y)$ of $G_2)$ for $\langle y, D, L\rangle \in$ ($ES2$ of $G_2$)". If $v = a_l$, $i \geq l$ and $a_l$ appears on some $S - \cdots - a_i - \cdots - D \subseteq ([ES1_{sub}]_S^D$ of $G)$, then there exists some $P = S - \cdots - v_1 - \cdots - t_1 \subseteq E_2$ (where $t_1$ lies at stage $L-2$) and $P[1{:}L-2] \nsubseteq (\lambda_{sub}(y)$ of $G_2)$ for $\langle y, D, L\rangle \in$ $(ES2$ of $G_2)$.

Since only $Pspare_{\hat{x}_l} =_{def} \left(\hat{x}_l - \omega_{l+1}^{\hat{x}_l} - \cdots - \omega_{L-1}^{\hat{x}_l} - D\right) = X^{\langle e|i|w\rangle}[l+1{:}L-2] \cup \left(x_{L-2}^{(e|i|w)} - \omega_{L-1}^{\hat{x}_l} - D\right) = \left(\hat{x}_l - x_{l+1}^{(e|i|w)} - \cdots - x_{L-2}^{(e|i|w)}\right) \cup \left(x_{L-2}^{(e|i|w)} - \omega_{L-1}^{\hat{x}_l} - D\right)$ in $G_2$ is changed to contain $T[1{:}L-1]$ and we only expand some labels in $G_2$, Remark 1, Remark 2 and Remark 3 can be proved for this $G_2$. This is a new $G_2$ different from the one that we defined earlier. Temporarily, in the following several lines, we still refer to it as $G_2$.

Then a contradiction will arise. According to (H1), we have $\left(\forall\langle y, D, L\rangle \in ([ES1_{sub}]_S^D \text{ of } G_2)\right)\left(\exists\, \sigma - \text{path } S - \cdots - y - D \subseteq (ES1_{sub} \text{ of } G_2)\right)$ (note that $y$ is some "$x_{L-1}^{(e|i|w)}$" here). Hence a $\sigma$-path exists in $G_2$, which is contained in $(ES1_{sub}$ of $G_2)$ and which contains the peer of the said $S - a_1 - \cdots - a_i$ in $G$. This peer essentially contains $S - a_1 - \cdots - a_i$. Then it can be inferred that there exists a corresponding $\sigma$-path $SP = S - a_1 - \cdots - a_i - \cdots - w - D \subseteq$ $(ES1_{sub}$ of $G)$ . Since $(ES1_{sub}$ of $G)$ satisfies criteria (i),(ii),(iii),(iv),(v), $S - a_1 - \cdots - a_i \subseteq SP[1{:}L-1] \subseteq$ $(\lambda_{sub}(w)$ of $G)$. Criterion (v) is violated. No such $(ES1_{sub}$ of $G)$ satisfying criteria (i),(ii),(iii),(iv),(v) exists.

**Case 1b.** $\left|([ES1_{sub}]_S^D[L{:}L] \text{ of } G)\right| = 1$.

If the unique path $P = S - \cdots - D \subseteq ([ES1_{sub}]_S^D$ of $G)$ is a $\sigma$-path, Claim 5 is proven. If $P$ is not a $\sigma$-path, after splitting, we must have two of those "$\langle x_{L-1}^{(e|i|w)}, D, L\rangle$" ( $i \in \{1,2\}$ ) in $(ES2$ of $G_2)$ at least. All the related "$\left(\lambda_{sub}\left(x_{L-1}^{(e|i|w)}\right) \text{ of } G_2\right)$" contain $P'[1{:}L-2]$ (where $P' = P[1{:}l-1] \cup (u_1 - v_1 - \cdots - D)$) when they are united together, but no one individually contains the whole $P'[1{:}L-2]$. Then the subsequent discussion can totally shift to the above discussion of **Case 1a**. We will encounter a contradiction again.

Summarizing the above discussions of Case 1a and Case 1b, $([ES1_{sub}]_S^D$ of $G)$ can only contain a unique path and the path must be a σ-path. Hence, $\left(\forall\langle w, D, L\rangle \in ([ES1_{sub}]_S^D \text{ of } G)\right)\left(\exists\, \sigma - \text{path } S - \cdots - w - D \subseteq (ES1_{sub} \text{ of } G)\right)$.

**Case 2.** $\begin{Bmatrix}\langle u_1, v, l\rangle, \\ \langle u_2, v, l\rangle\end{Bmatrix} \cap ([ES1_{sub}]_S^D \text{ of } G) \in \begin{Bmatrix}\emptyset, \\ \{\langle u_1, v, l\rangle | (R(\langle u_2, v, l\rangle) \text{ of } G) = \emptyset\}\end{Bmatrix}$.

Note that $([ES1_{sub}]_S^D$ of $G)$ cannot contain both in-degrees of $v$ by criterion (iv), and the condition $(R(\langle u_2, v, l\rangle)$ of $G) = \emptyset$ implies no $\langle x_{L-1}^{(e|2|w)}, D, L\rangle \in (ES1$ of $G_2)$.

In Case 2, $([ES1_{sub}]_S^D[L{:}L]$ of $G_2) = (ES2$ of $G_2)$ already holds, so no compensation is needed. Note that $([ES1_{sub}]_S^D$ of $G_2)$ now keeps naturally the same with $([ES1_{sub}]_S^D$ of $G)$—for $\langle w, D, L\rangle \in ([ES1_{sub}]_S^D$ of $G)$, there always exists some $\langle x_{L-1}^{(e|i|w)}, D, L\rangle \in ([ES1_{sub}]_S^D$ of $G_2)$. Since $([ES1_{sub}]_S^D[L{:}L]$ of $G_2) = (ES2$ of $G_2)$, $(ES1_{sub}$ of $G_2)$ satisfies criteria (i),(ii),(iii),(iv),(v) (by Remark 4) and $f(G_2) < f(G)$ , we know that $\left(\forall\langle y, D, L\rangle \in ([ES1_{sub}]_S^D \text{ of } G_2)\right)\left(\exists\, \sigma - \text{path } S - \cdots - y - D \subseteq (ES1_{sub} \text{ of } G_2)\right)$ by (H1). It follows that $\left(\forall\langle w, D, L\rangle \in ([ES1_{sub}]_S^D \text{ of } G)\right)\left(\exists\, \sigma - \text{path } S - \cdots - w - D \subseteq (ES1_{sub} \text{ of } G)\right)$.

We thus complete the proof of Claim 5.

The above Claim 1,2,3,4,5 conclude the proof of Lemma 2.

## A.7 Proof of Theorem 6

*Proof.* Let $G =< V, E, S, D, L, \lambda >$ be the multi-stage graph in the $2 - \text{MSP}$ input to the ZH algorithm.

Pick any $\langle y, D, L\rangle \in ES1[L{:}L]$, we can choose $ES2 = \{\langle y, D, L\rangle\}$. Since $[R(\langle x, y, L-1\rangle) \cap ES1]_y^D \neq \emptyset$ for an arbitrary $\langle x, y, L-1\rangle \in ES1$, then $\langle y, D, L\rangle \in R(\langle x, y, L-1\rangle)$ and there exists $P = S - a - \cdots - y - D \subseteq A$ for the set $A$ computed when deciding $\langle y, D, L\rangle \in R(\langle x, y, L-1\rangle)$, we can choose $\lambda_{sub}(y) = S - a - \cdots - y \subseteq \lambda(y)[1{:}1] \cup \{e \in \lambda(y)[2{:}L] | \langle y, D, L\rangle \in R(e)\}$.

Then, we obtain $ES1_{sub} = ES2 \cup (\lambda_{sub}(y) \cap ES1) \subseteq ES1$.

Further, we can obtain that $[ES1_{sub}]_S^D[L{:}L] = ES2 \neq \emptyset$, when noting that $A \subseteq ES1$.

$ES1_{sub}$ fulfills the criteria (i),(ii),(iv). $ES1_{sub}$ cannot obey criterion (iii), because we can assume $G$ has no $\sigma$-paths. $ES1_{sub}$ obeys criterion (v), since $|ES2| = 1$.

Then, $G$ must contain a $\sigma$-path $SP \subseteq ES1_{sub}$ claimed by the PA using the αβ lemma.☐

## A.8 Supplementary materials for the proof of Lemma 2

### *A.8.1 The renaming rules and the "transit" technique for $(R(E)\ of\ G_1)$*

For every $\{\langle r, s, k\rangle, \langle o, p, h\rangle\} \subseteq E$ ($1 \leq k < h \leq L$), if $\langle o, p, h\rangle \in (R(\langle r, s, k\rangle) \text{ of } G)$ (where $(R(\langle r, s, k\rangle) \text{ of } G) \in (R(E) \text{ of } G)$), there must exist $\{e_1, e_2\} \subseteq E_1$, such that $e_2 \in (R(e_1) \text{ of } G_1)$ (where $(R(e_1) \text{ of } G_1) \in (R(E) \text{ of } G_1)$). This should hold for both initial and constrained $\rho$-path edge-sets. Here are the detailed renaming rules for the above $e_1$,$e_2$.

***The renaming rules for $(R(E)$ of $G_1)$:***

**Case 1.** ($\langle o, p, h\rangle \notin \left((u_i - v - \cdots - D) \text{ of } G\right), i \in \{1,2\}$):

$e_1 = \langle r, s, k\rangle, e_2 = \langle o, p, h\rangle$.

**Case 2.** ($\langle o, p, h\rangle \in \left((u_i - v - \cdots - D) \text{ of } G\right), i \in \{1,2\}$):

2.1. $\langle o, p, h\rangle = (\langle u_i, v, l\rangle \text{ of } G)$:

$e_1 = \langle r, s, k\rangle, e_2 = \langle u_i, v_i, l\rangle$.

2.2. ($\langle o, p, h\rangle \in \left((v - \cdots - D) \text{ of } G\right)$):

2.2.1. $k < l$:

$e_1 = \langle r, s, k\rangle, e_2 \in \left((v_j - \cdots - D)[h{:}h] \text{ of } G_1\right)$

(such that $\langle u_j, v, l\rangle \in (R(\langle r, s, k\rangle) \text{ of } G), j \in \{1,2\}$).[4]

[4] If $\{\langle u_1, v, l\rangle, \langle u_2, v, l\rangle\} \subseteq (R(\langle r, s, k\rangle) \text{ of } G)$, there are two edges (i.e., the ones in $\cup_{j\in\{1,2\}}\left((v_j - \cdots - D)[h{:}h]\right)$) in $G_1$ each corresponding to the $\langle o, p, h\rangle$ in $G$. It's similar for the other cases.

2.2.2. $k = l$:

$e_1 = \langle u_j, v_j, l \rangle, e_2 \in \left( (v_j - \cdots - D)[h{:}\,h] \text{ of } G_1 \right)$

(such that $\langle r, s, k \rangle = \langle u_j, v, l \rangle, j \in \{1,2\}$).

2.2.3. $k > l$:

$e_1 \in \left( (v_j - \cdots - D)[k{:}\,k] \text{ of } G_1 \right), e_2 \in \left( (v_j - \cdots - D)[h{:}\,h] \text{ of } G_1 \right)$

(such that $\langle o, p, h \rangle \in \left( R(\langle u_j, v, l \rangle) \text{ of } G \right), j \in \{1,2\}$).

The key technique used here—to clarify the discussion of the many edges and paths involved in the computations of Operator 2,3,4—is a **"transit"** between $(R(e) \text{ of } G)$ and $(R(e) \text{ of } G_1)$ on the multi-in-degree vertex $v$ in $G$:

1. Once $\langle u_i, v, l \rangle \in (R(e) \text{ of } G)$ ( $i \in \{1,2\}$ ), then $\langle u_i, v_i, l \rangle \in (R(e) \text{ of } G_1)$ (by the radical expansion on $(\lambda(v_i) \text{ of } G_1)$);
2. Then straightly, we can have $v_i - \cdots - w_i - D \subseteq (R(e) \text{ of } G_1)$ (by the radical expansion on $(\lambda(t_i) \text{ of } G_1)$,...,$(\lambda(w_i) \text{ of } G_1)$).

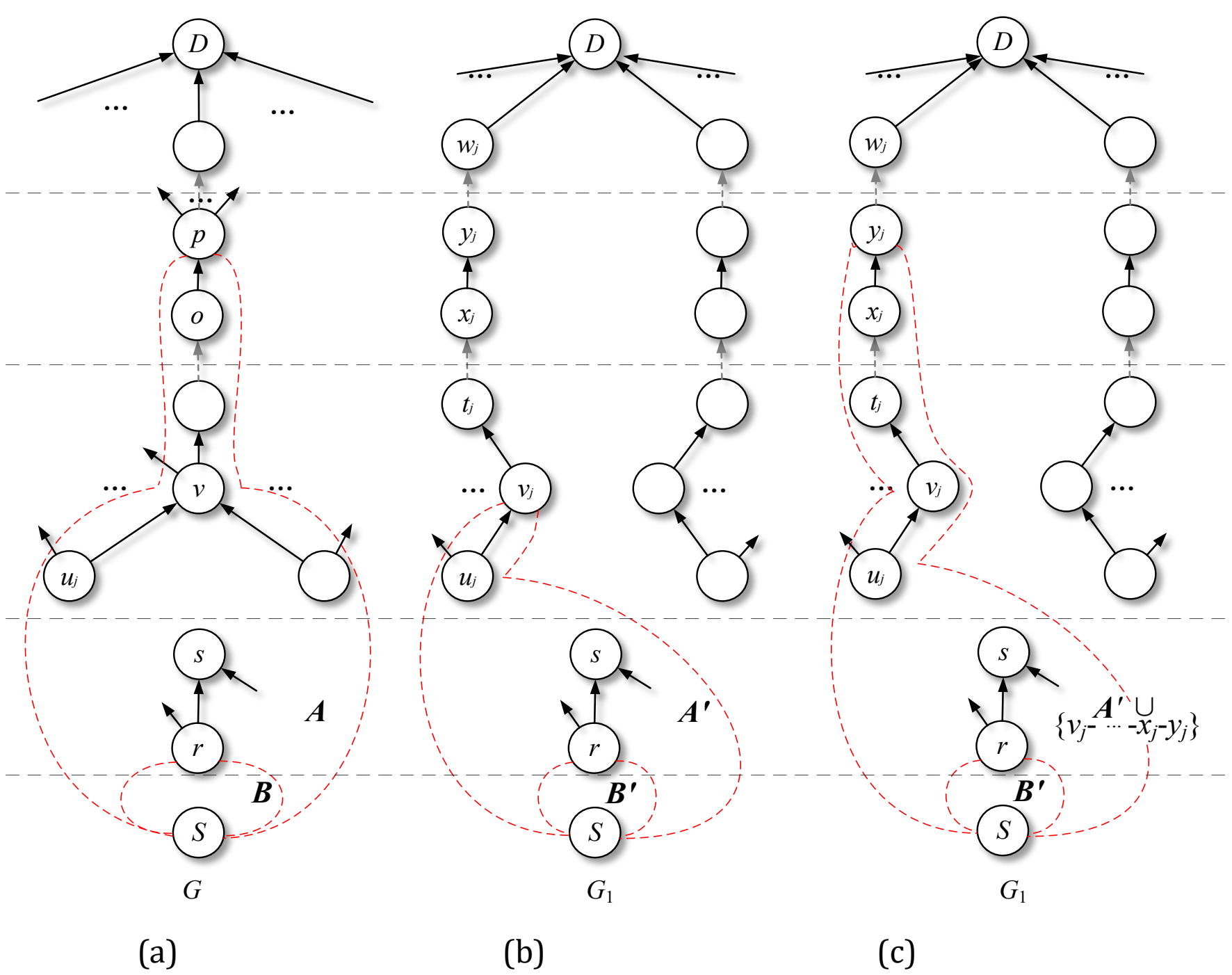

Figure 11: A typical case (Case 2.2.1) of the renaming rules for $(\mathrm{R(E)} \text{ of } \mathrm{G}_1)$

The "transit" is the direct consequence of the radical expansion. It ensures that, the initial $\rho$-paths (by Operator 2) and constrained $\rho$-paths (by Operator 4) of each edge in $G$ are "naturally preserved" in $G_1$ despite the "split" of $v$, when computing $(R_0(e) \text{ of } G_1)$ by the definition of Operator 2 and when computing $(R(e) \text{ of } G_1)$ by the definition of Operator 4. It saves the heavy efforts otherwise required to dive into the details of the operators, especially the convoluted Operator 4.

Now, let's utilize this technique to explain the above naming rules. We felt it unnecessary to exquisitely use inductive proof here, under the technique of "transit" and for the brevity of the proof; although it might will be slightly more rigorous in that manner. Besides, the ZH algorithm already naturally provides a inductive framework with the iterative steps of its pseudo-code. Throughout the iterative steps, the result to be proved is consistently guaranteed by the "transit" technique.

For $\langle o,p,h\rangle \in (R(\langle r,s,k\rangle) \text{ of } G)$, w.l.o.g., let's suppose $k < l < h$ (i.e., Case 1 or Case 2.2.1, other cases are analogous):

1. When $k = 1$, things become trivial, since $(R(\langle r,s,k\rangle) \text{ of } G_1) = (R_0(\langle r,s,k\rangle) \text{ of } G_1)$ (recall the ZH algorithm skips the first stage). Thus, assume $k > 1$ hereinafter.
2. If $\langle o,p,h\rangle \notin [E]_v^D$, then $\{\langle r,s,k\rangle, \langle o,p,h\rangle\} \subseteq E_1$. By the construction of $G_1$, $\langle o,p,h\rangle \in (R(\langle r,s,k\rangle) \text{ of } G_1)$.
3. Otherwise if $\langle o,p,h\rangle \in [E]_v^D$, by the definition of Operator 4, the set ($\boldsymbol{A}$ of $G$) involved in the computation for deciding "$\langle o,p,h\rangle \in (R(\langle r,s,k\rangle) \text{ of } G)$" is non-empty (see Figure 11(a)). Since $\{\langle u_i,v,l\rangle, \langle o,p,h\rangle\} \subseteq (\boldsymbol{A} \text{ of } G)$ ($i \in I, I \subseteq \{1,2\}$), then by the definition of Operator 3 for computing ($\boldsymbol{A}$ of $G$), $\langle u_j,v,l\rangle \in (R(\langle r,s,k\rangle) \text{ of } G)$ ($j \in J, J \subseteq I$).
4. By step (1) of the "transit" technique, we then have $\langle u_j,v_j,l\rangle \in (R(\langle r,s,k\rangle) \text{ of } G_1)$ (see Figure 11(b)). By step (2) of the "transit" technique, we further have $v_j - \cdots - w_j - D \subseteq (R(\langle r,s,k\rangle) \text{ of } G_1)$ (see Figure 11(c)).

Thus, there exists $\langle r,s,k\rangle$ and $e_2 \in \left(v_j - \cdots - D\right)[h{:}\,h]$ in $G_1$, such that $e_2 \in (R(\langle r,s,k\rangle) \text{ of } G_1)$.

#### *A.8.2 The computation of* $\left(\chi_{R(E)}^D(ES_temp) \text{ of } G_1\right)$

For the chosen ($ES_temp$ of $G_1$), we can obtain that $\left(\chi_{R(E)}^D(ES_temp) \text{ of } G_1\right) \neq \emptyset$. The argument for it is as follows.

Firstly, since ($\boldsymbol{A}$ of $G$) $\neq \emptyset$ and ($\boldsymbol{B}$ of $G$) $\neq \emptyset$, ($\boldsymbol{A}$ of $G$) must contain $\langle u_i,v,l\rangle$ for $i \in I$ ($I \subseteq \{1,2\}$), and then we must have $u_i - v_i - \cdots - w_i - D \subseteq (ES_temp \text{ of } G_1)$.

Further, we are to show that there exists a non-empty edge set ($\boldsymbol{A}'$ of $G_1$) in $G_1$ computed essentially the same as the set ($\boldsymbol{A}$ of $G$) in $G$, where

$$(\boldsymbol{A}' \text{ of } G_1) = \\ (\boldsymbol{A}[1{:}\,l-1] \text{ of } G) \cup \left\{ e' \,\middle|\, \begin{matrix} e' \in u_i - v_i - \cdots - w_i - D \subseteq E_1, \\ \langle u_i,v,l\rangle \in (\boldsymbol{A} \text{ of } G), i \in \{1,2\} \end{matrix} \right\}. \tag{18}$$

- Firstly, by the radical expansion, $u_i - v_i - t_i - \cdots - w_i - D$ is a $\omega$-path in $G_1$ and hence $\left([R(\langle c,d,k\rangle) \cap \boldsymbol{A}']_d^D \text{ of } G_1\right) \neq \emptyset$ for each $\langle c,d,k\rangle \in (\boldsymbol{A}' \text{ of } G_1)$ ($l \leq k < L$).
- Secondly, for each $\langle c,d,k\rangle \in (\boldsymbol{A}[1{:}\,l-1] \text{ of } G) \subseteq (\boldsymbol{A}' \text{ of } G_1)$ ($k < l$), since $\langle c,d,k\rangle \in (\boldsymbol{A} \text{ of } G) = \left(\chi_{R(E)}^D(\boldsymbol{A}) \text{ of } G\right)$, then each $P_1 = d - \cdots - D \subseteq \left(\left[R(\langle c,d,k\rangle) \cap \chi_{R(E)}^D(\boldsymbol{A})\right]_d^D \text{ of } G\right)$ must traverse $\langle u_i,v,l\rangle$ for some $i = 1$ or 2. Hence, $\langle u_i,v_i,l\rangle \in (R(\langle c,d,k\rangle) \text{ of } G_1)$ (by the radical expansion of ($\lambda(v_i)$ of $G_1$)) and further $u_i - v_i - t_i - \cdots - w_i - D \subseteq (R(\langle c,d,k\rangle) \text{ of } G_1)$ (by the radical expansion of ($\lambda(t_i)$ of $G_1$),…,($\lambda(w_i)$ of $G_1$)). [Also see the renaming rules and the "transit" technique discussed in Appendix A.8.1 for ($R(E)$ of $G_1$) if needed.) Thus, $P_2 = [P_1]_d^{u_i} \cup (u_i - v_i - \cdots - w_i - D) \subseteq$ $\left([R(\langle c,d,k\rangle) \cap \boldsymbol{A}']_d^D \text{ of } G_1\right)$.

- Subsequently, by the definition of Operator 3, we have ($\boldsymbol{A}'$ of $G_1$) $= \left(\chi_{R(E)}^D(\boldsymbol{A}') \text{ of } G_1\right) \neq \emptyset$.

Finally, we now intent to show that the aforementioned edge set ($\boldsymbol{A}'$ of $G_1$) will still be non-empty when compacted by Operator 3, even all its edges at stage $h$ are removed except $\langle a,b,h\rangle$; in other words, $\left(\chi_{R(E)}^D(ES_temp) \text{ of } G_1\right) \neq \emptyset$. To show this, for each $\hat{e} = \langle c,d,k\rangle \in$ ($ES_temp$ of $G_1$), consider the following cases (akin to the discussion happened during the proof of Claim 2 for $G_1$):

(i) $h < k < L$. Straightforwardly, $\langle c,d,k\rangle \in$ ($ES_temp$ of $G_1$) implies $\langle c,d,k\rangle \in$ ($\boldsymbol{A}'$ of $G_1$). Hence, $\left([R(\langle c,d,k\rangle) \cap ES_temp]_d^D \text{ of } G_1\right) = \left([R(\langle c,d,k\rangle) \cap \boldsymbol{A}']_d^D \text{ of } G_1\right) \neq \emptyset$.

(ii) $k = h$. Then, we have $\langle c,d,k\rangle = \langle a,b,h\rangle$. Analogous to (i), we can obtain that $\left([R(\langle c,d,k\rangle) \cap ES_temp]_d^D \text{ of } G_1\right) \neq \emptyset$.

(iii) $1 \leq k < h$. Since ($\boldsymbol{B}$ of $G$) $= \left(\chi_{R(E)}^a(\boldsymbol{B}) \text{ of } G\right) \subseteq$ ($ES_temp$ of $G_1$) is also a compacted edge set by Operator 3, it is sufficient to only consider those $\langle c,d,k\rangle \in$ ($\boldsymbol{B}$ of $G$) for $\left(\chi_{R(E)}^D(ES_temp) \text{ of } G_1\right)$ (see Figure 8). Since for the arbitrary enumerated $\langle c,d,k\rangle \in \left(\chi_{R(E)}^a(\boldsymbol{B}) \text{ of } G\right) \subseteq$ ($ES_temp$ of $G_1$), there exists $P_1 = d - \cdots - a - b - \cdots - u_i - v - \cdots - D \subseteq \left((R(\langle c,d,k\rangle) \cap \boldsymbol{A}) \text{ of } G\right)$ (for some $i \in \{1,2\}$), then $[P_1]_d^{u_i} \cup \langle u_i, v_i, l\rangle \subseteq$ ($R(\langle c,d,k\rangle)$ of $G_1$) (by the radical expansion on ($\lambda(v_i)$ of $G_1$)) and $P_2 = [P_1]_d^{u_i} \cup (u_i - v_i - t_i - \cdots - w_i - D) \subseteq \left((R(\langle c,d,k\rangle) \cap ES_temp) \text{ of } G_1\right)$ (by the radical expansion on ($\lambda(t_i)$ of $G_1$),…,($\lambda(w_i)$ of $G_1$)). (Also see the renaming rules and the "transit" technique discussed in Appendix A.8.1 for ($R(E)$ of $G_1$) if needed.)

Therefore, by the definition of Operator 3, we obtain that $\left(\chi_{R(E)}^D(ES_temp) \text{ of } G_1\right) \neq \emptyset$.

These two pages do not constitute part of the main paper. They serve solely as a recommended list of queries for AI engagement. Additionally, they can provide an overview of the organizational framework of the main paper and clarify the interrelationships between its conclusions.

The AI query list (you may add more questions as needed)

1. Is Theorem 1 ($NP$-completeness) correct? That is, is $2-MSP$ an $NP$-complete problem?

2. Is Theorem 2 correct? That is, is the result of Operator 3 is uniquely determined, regardless of the order in which edges are chosen for pruning during the iteration?

3. Is Theorem 3 correct? That is, is the result of Operator 4 is uniquely determined, regardless of the order in which edges are chosen for pruning during the iteration?

4. Are the complexities of operator 1, operator 2, operator 3, and operator 4 polynomial functions of $|E|$?

5. Is Theorem 4 correct? That is, is the cost of the ZH algorithm never more than $O(|E|^{10})$?

6. Is Theorem 5 correct? That is, if $G$ contains a $\sigma$-path, then, is the compact kernel of $G$ non- empty?

7. Is Lemma 1 correct?

8. Is Claim 1 in the proof of Lemma 2 correct? That is, does $f(G_1) < f(G)$?

9. Is Claim 2 in the proof of Lemma 2 correct?

10. Is Claim 3 in the proof of Lemma 2 correct? That is, does $G_1$ satisfy Definition 3(b)?

11. In the proof of Claim 4a of Lemma 2, the paper needs to define ($ES1_{sub}$ of $G_1$) and ($ES2$ of $G_1$). Is the defined ($ES1_{sub}$ of $G_1$) satisfies criteria (i),(ii),(iii),(iv) and (v)? Does ($[ES1_{sub}]_S^D[L:L]$ of $G_1$) $=$ ($ES2$ of $G_1$) for the defined ($ES1_{sub}$ of $G_1$) and ($ES2$ of $G_1$)?

12. Is Claim 4a correct?

13. Is Claim 4b correct?

14. Is Claim 4c correct?

15. Is Claim 4 correct?

16. In the proof of Claim 5 of Lemma2, for the defined $G_2$, is Remark 1 correct? That is, does $f(G_2) < f(G)$?

17. In the proof of Claim 5 of Lemma2, for the defined $G_2$, is Remark 2 correct?

18. In the proof of Claim 5 of Lemma2, for the defined $G_2$, is Remark 3 correct? That is, is $G_2$ a $2-\text{MSP}$ problem?

19. In the proof of Claim 5 of Lemma2, for the defined $G_2$, is Remark 4 correct? That is, does ($ES1_{sub}$ of $G_2$) satisfy the criteria (i),(ii),(iii),(iv),(v) defined in the $PA$ algorithm?

20. In the proof of Claim 5 of Lemma2, in Case 1a, based on $G_2$ and ($ES1_{sub}$ of $G_2$), a new $G_2$ and new ($ES1_{sub}$ of $G_2$) are defined. To do so, the paper adds a new $\sigma$-path in $G_2$, namely $T$ in the paper. Can that $T$ be constructed? Does ($[ES1_{sub}]_S^D[L:L]$ of $G_2$) $=$ ($ES2$ of $G_2$) for this new ($ES1_{sub}$ of $G_2$)?

21. In the proof of Claim 5 of Lemma2, in Case 1a, a new $G_2$ and new ($ES1_{sub}$ of $G_2$) are defined. For the new $G_2$, can Remark 1, Remark 2 and Remark 3 be proved similarly?

22. In the proof of Claim 5 of Lemma2, in Case 1a, a new $G_2$ and new ($ES1_{sub}$ of $G_2$) are defined. Is this new ($ES1_{sub}$ of $G_2$) satisfying criteria (i),(ii),(iii),(iv),(v) defined in the $PA$ algorithm?

23. In the proof of Claim 5 of Lemma2, within Case 1a, the paper refutes the statement $\left|([ES1_{sub}]_S^D[L{:}\,L] \text{ of } G)\right| > 1$ by means of mathematical induction. Is the reasoning underlying this refutation correct?

24. In Case 1b, suppose $\left|([ES1_{sub}]_S^D[L{:}\,L] \text{ of } G)\right| = 1$, and the unique path $P = S - \cdots - D \subseteq ([ES1_{sub}]_S^D \text{of } G)$ is not a $\sigma$-path, the paper refutes this scenario via mathematical induction. Is the reasoning for the refutation correct?

25. In the proof of Claim 5 of Lemma2(Case 1), the paper infers via mathematical induction that: $\left(\forall\langle w, D, L\rangle \in ([ES1_{sub}]_S^D \text{ of } G)\right)\left(\exists\, \sigma - \text{path } S - \cdots - w - D \subseteq (ES1_{sub} \text{ of } G)\right)$. Is the inferred conclusion valid under the specific conditions inherent to Case 1?

26. In the proof of Claim 5 of Lemma2, within Case 2, the paper infers via mathematical induction that: $\left(\forall\langle w, D, L\rangle \in ([ES1_{sub}]_S^D \text{ of } G)\right)\left(\exists\, \sigma - \text{path } S - \cdots - w - D \subseteq (ES1_{sub} \text{ of } G)\right)$. Is the inferred conclusion valid under the specific conditions inherent to Case 2?

27. Is Claim 5 correct?

28. Is Lemma 2 correct?

29. Is the αβ lemma (Summarizing Lemma 1 and 2) correct?

30. Is Theorem 6 (the sufficiency) correct?

31. Then, based on your reading and all the above discussion, is Theorem 7 (***NP*** = ***P***) correct? That is, is ZH algorithm a polynomial-time algorithm for solving the $2-$MSP problem?